\documentclass[11pt]{article}
\usepackage[T1]{fontenc}
\usepackage[square,numbers]{natbib}
\usepackage{graphics}
\usepackage[a4paper, total={6.8in, 10.1in}]{geometry}
\renewcommand{\baselinestretch}{1.25}
\usepackage[english]{babel}
\usepackage{graphicx}
\usepackage{amsfonts}
\usepackage{mathptm}
\usepackage{pstricks}
\usepackage{pst-node}
\usepackage{amsmath}
\usepackage{ulem}
\usepackage{authblk}
\usepackage{url}
\usepackage{booktabs}
\usepackage{subcaption}
\usepackage{multirow}
\usepackage{rotating}

\usepackage{hyperref}

\newcommand{\SP}[1]{{\color{black} #1}}

\providecommand{\keywords}[1]
{
  \textbf{\textit{Keywords---}} #1
}

\title{\vspace{2cm} \Large \textbf{Modeling and solving a vehicle-sharing problem considering multiple alternative modes of transport}}
\bigskip
\author[a,b]{\bf Miriam Enzi \footnote{Corresponding author.}}
\author[a]{\bf Sophie N. Parragh}
\author[c]{\bf David Pisinger}
\bigskip
\affil[a]{\small Institute of Production and Logistics Management, Johannes Kepler University Linz \protect\\
Altenberger Straße 69, 4040 Linz, Austria \protect\\
\texttt{\{miriam.enzi,sophie.parragh\}@jku.at}
\medskip }
\affil[b]{Center for Mobility Systems, AIT Austrian Institute of Technology \protect\\
Giefinggasse 4, 1210 Vienna, Austria \protect
\medskip }
\affil[c]{Department of Management Engineering, Technical University of Denmark \protect\\
Akademivej Building 358, 2800 Kgs. Lyngby, Denmark \protect\\
\texttt{pisinger@man.dtu.dk}
\medskip }

\newcommand{\dis}            {\displaystyle}

\newcommand{\minflowcar}    {{\textit{VShP-1T:car}}}
\newcommand{\minflowecar}    {{\textit{VShP-1T:ecar}}}
\newcommand{\multcomm}    {{\textit{VShP-xT}}}

\newcommand{\singlesharing}    {{\textit{VShP-1T}}}
\newcommand{\multisharing}    {{\textit{VShP-xT}}}
\newcommand{\car} {{car}}
\newcommand{\ecar} {{ecar}}
\newcommand{\mix} {{mix}}

\begin{document}
\renewcommand{\baselinestretch}{1.0}

\begin{titlepage}
\maketitle
\thispagestyle{empty}


\begin{abstract}
Motivated by the change in mobility patterns, we present a scheduling approach for a vehicle-sharing problem, \SP{considering several alternative modes of transport,} from a company viewpoint with centralized planning. We consider vehicle-sharing in a company having one or more depots and a fixed number of users, i.e. employees. The users have appointments with a fixed location and fixed start and end times. A vehicle must be used for a full trip of a user from depot to depot. We aim at assigning vehicles to user trips so as to maximize savings compared to other modes of transport.  

We first consider that only one type of vehicle is used, and second that multiple vehicle types can be \SP{used}. For the first case, we show that the vehicle-sharing problem can be formulated as a minimum-cost flow problem. Secondly, if multiple types of vehicles are available the problem can be formulated as a multi-commodity flow problem. These formulations make the problem applicable in daily operations due to efficient solution methods.

We provide a comprehensive computational study for both cases \SP{on} 
instances based on demographic, spatial, and economic data of Vienna. We show that our formulations for this problem solve \SP{these} 
instances in a few seconds, which makes them usable in an online booking system. In the analysis we discuss \SP{different potential settings. We }  
study the optimal composition of a shared fleet, restricted sets of modes of transport, and variations of the objective function.

\end{abstract}

\bigskip

\end{titlepage}
\newpage



\section{Introduction} 

Mobility -- how we use it and see it -- is changing. People tend to be mobile rather than owning cars. ''Mobility as a Service'' (MaaS) \citep{Mulley2018} has emerged as a widely known and used term. This change is supported by novel mobility concepts: \SP{ in the private sectors and in the area of corporate mobility}. Companies are trying to change their view on their corporate mobility by switching from individually assigned cars towards MaaS for their employees, and give incentives to use (a combination of) ''greener'' modes of transport to avoid pollution and congestion problems. Having shared mobility within a company (or any other closed group of users such as, e.g., home communities) will be increasingly important in future mobility settings. 

In this work we study \SP{a corporate mobility problem} 
and report results for 
\SP{instances generated during }
an applied research project with several company partners (http://www.seamless-project.at). It is a vehicle-sharing problem in a company, having one or more offices (depots), from which the employees (users) have to visit various customers during office hours (e.g. for business meetings). Each visit (task) involves one specific user and has a fixed time, which gives us a fixed sequence of tasks. A trip covers the fixed sequence of tasks of one user, starting at a depot and terminating at the same location or in another depot of the company. Thus a trip contains several stops and it starts and ends at a predefined (but possibly different) depot.
The company operates a pool of shared vehicles of a fixed size and provides possibilities to use other modes of transport (MOT), such as bikes, taxis, \SP{public transport} or walk, \SP{which are not shared.}
Different to most other vehicle-sharing problems, we study the problem from a company viewpoint with centralized planning. Thus, we minimize a company's expenses and do not focus on individual goals.

\begin{figure}[t]
    \begin{subfigure}[c]{\textwidth}
    \centering
      \includegraphics[width=0.9\textwidth]{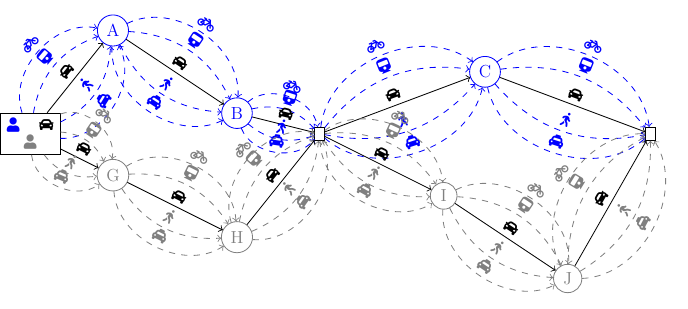}
    \subcaption{Users, modes of transport, and task sequences in a time-space graph \label{subfig:tasks}} 
    \end{subfigure}
   \begin{subfigure}[c]{\textwidth}
   \centering
    \includegraphics[width=0.9\textwidth]{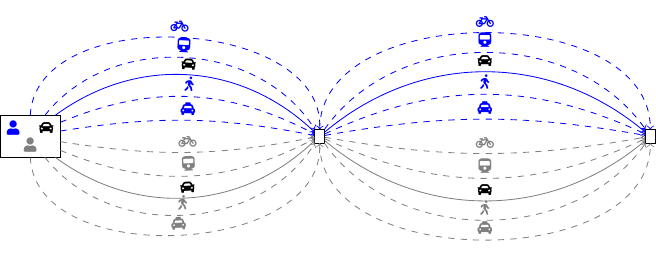}
     \subcaption{Users, modes of transport, and trips  in a time-space graph \label{subfig:trips}} 
    \end{subfigure}
    \caption{An example for the considered car sharing problem with two users and one shared mode of transport (a vehicle) }
    \label{fig:example}
\end{figure}

\SP{In Figure~\ref{fig:example}, we provide an illustrative example with two users, one office location and one shared vehicle. The first user has to travel to complete tasks A, B, and C and the second user has to cover tasks G, H, I, and J (see Figure~\ref{subfig:tasks}). The timing of the tasks is fixed, and if a return to the office is not possible between two tasks, they are modeled on the same trip and are replaced by an arc as illustrated in Figure~\ref{subfig:trips}.}

The aim is to assign user-trips to the available vehicles, e.g., shared cars, so as to maximize the savings obtained when using a vehicle instead of other MOTs. The costs of transportation include distance cost, but also hourly wages of employees in order to properly reflect the trade-off between fast (but expensive) and possibly slower (but cheaper) modes of transport, such as public transportation or bikes. We note that cars may not always be the fastest (or cheapest) MOT. Furthermore, we also include emission cost to strengthen the use of environmentally friendly MOTs. 

Many formulations studying car-sharing systems are based on time-space networks such as, e.g., \citet{Correia2012}, considering depot locations in the context of one-way car-sharing.
\citet{Brandstaetter2020} model the movement of cars in a electric car-sharing system as a multi-commodity flow problem.
\citet{Zhang2019} work on vehicle-to-trips assignment and relay decisions in one-way car-sharing systems with electric vehicles. They model the problem in a single-commodity network flow model and develop a heuristic thereof.
In \citet{Enzi2020} both car-sharing and ride-sharing are considered simultaneously and in combination with alternative MOTs. The first step of the auxiliary graph transformation of \citet{Enzi2020} is similar to the presented graph in this work. However, they extend the graph by duplicating trips including ride-sharing and solve the car- and ride-sharing problem by a kind of column generation algorithm, assigning cars to trips. \SP{A time-dependent problem version considering a company and a user objective concurrently is addressed in \citet{enzi2021bi}.}
Detailed surveys on car-sharing are provided by \citet{Jor2013}, \citet{Brandstatter2016} and \citet{Laporte2018}.

In our paper, we rely on the modeling of two well-known network problems, namely the minimum-cost flow problem and the multi-commodity flow problem. 
Even though similar modeling approaches have been applied \SP{in the context of other car-sharing problems, we give a theoretical contribution outlining that also the  sharing systems considered in this paper, including multiple alternative modes of transport, in a company setting} 
can be modeled using these well-known formulations.
We formulate the case where only one type of vehicle is shared as a minimum-cost flow problem \citep{Ahuja2001}. If more than one type of vehicles is shared, we base the formulation on the multi-commodity flow problem \citep{Barnhart2001}. Note that, even though we will mainly base our examples and results on cars, this problem can easily incorporate other shared vehicles, such as bikes or scooters.


\paragraph{\textbf{Contribution and outline}}

The contributions of this paper are as follows: we introduce and model a corporate vehicle-sharing problem with predetermined trips, \SP{considering multiple different MOTs}, using the well-known minimum-cost flow and multi-commodity flow formulations, which can be solved efficiently and thus used in an online operational setting within a company with centralized planning. Furthermore, we provide a detailed analysis with respect to the impact of using different kinds of shared vehicles, and provide insights into optimal fleet composition in a shared system. We also analyze the number of trips per vehicle during a day and the disadvantage (from a cost-perspective) when giving the opportunity to restrict the set of available MOTs per user/trip. We compare the case where no sharing is allowed with our introduced sharing systems. Finally, we compare the outcomes of different objective functions, where first we use a combination of operational distance cost and cost of time, and then consider time only.

The paper is organized as follows: We start by introducing our vehicle-sharing problem in Section~\ref{sec:carshare}. We first introduce the model with a single shared vehicle type, formulated as a minimum-cost flow problem in Section~\ref{sec:singlesharing}, followed by the model with multiple shared vehicle types formulated as a multi-commodity flow problem in Section~\ref{sec:multisharing}. In Section~\ref{sec:compresults} we summarize our analysis based on an extensive computational study and give managerial implications using instances based on demographic, spatial, and economic data of Vienna, Austria. We conclude this paper in Section~\ref{sec:concl}.

\section{A vehicle-sharing problem}\label{sec:carshare}

Formally, our vehicle-sharing problem can be formulated as follows:

We have a set of users $P$ that have to visit their scheduled meetings (tasks).
Each task is associated with a different location and has an associated fixed start time and duration. The user-to-task assignment is not interchangeable, resulting in a fixed sequence of tasks per user.
Every user $p \in P$ covers one or more trips $\pi$. A trip has an origin $o_{\pi}$ and destination $e_{\pi}$ whilst covering in between a fixed set of tasks. 
Moreover, we consider a set of modes of transport $K$ such as walk, bikes, public transportation (bus, train, metro), taxis and cars, where at least one MOT $k$ has a restricted capacity that is shared, e.g., cars. 
If a trip is started with one mode of transport, then the same mode should be used for the full trip.


\SP{Let $C^k_{\pi}$ denote the cost for covering trip $\pi$ with MOT $k$, then for } 
each trip $\pi$ let min$_{k \in K \setminus \{1\}} C^k_{\pi}$ be the cost of the cheapest mobility type excluding cars $k=1$ (assuming that we are sharing cars). Let $C^{1}_{\pi}$ be the cost of riding the same trip $\pi$ by car $k=1$. We then calculate the savings $s_{\pi} = C^{K \setminus \{1\}}_{\pi} - C^{1}_{\pi}$ of using a car compared to using the cheapest possible other mobility type. \SP{Savings may be negative in cases where the car is not the cheapest MOT to cover a given trip.}
Note that if traveling with a certain MOT is not possible, we impose a penalty and set $C_{\pi}^k = \infty$. 

Finally, we aim at assigning user-trips to the shared vehicles in the best possible way whilst maximizing savings obtained when using a car compared to the cheapest other mobility type. \SP{This allows us to consider only the shared MOTs with limited availability in our models while minimizing the cost of the entire system.}

We model the problems on a directed acyclic graph (DAG). 
Since a MOT must be used for the full trip, we do not model the tasks covered by a trip in the graph, and only consider nodes $o_{\pi}$ and $e_{\pi}$ for each trip $\pi$, which represent starting and ending points of a trip. The savings of the arc $(o_{\pi},e_{\pi})$ is $s_{\pi}$, as explained above. In order to connect the trips we insert additional arcs $(e_{\pi}, o_{\pi'})$ if trip $\pi_{\pi}$ has the same destination as trip $\pi_{\pi'}$ has origin, and the trip $\pi$ finishes before the trip $\pi'$. The savings of such an arc is $0$. \SP{ We denote the set of these arcs by $\mathcal{A}_{\pi}$}.

In the following, we introduce the modeling of the two cases presented in this paper. First, we introduce the modeling approach for the case where only one type of vehicle is shared and then solve it as a minimum-cost flow problem. Second, we present the formulation where multiple shared vehicle types can be employed. This is then modeled and solved as a multi-commodity flow problem.

\subsection{The vehicle-sharing problem with a single type of shared vehicle (\singlesharing)}
\label{sec:singlesharing}

For the  vehicle-sharing problem with a single type of shared vehicle (\singlesharing) we create a node $A_d$ for each depot $d \in D$ with a supply $\delta_d$ representing the number of available vehicles. Depots represent locations where the shared vehicles start and end, e.g. a company's offices. For each depot $d$ where the vehicles must be parked in the evening, we create a node $A'_d$ with a demand $\delta'_d$ equal to the number of requested vehicles at the end of the \SP{day}. Every node $A_d$ is connected to all nodes $o_{\pi}$ if trip ${\pi}$ starts in depot $d$. Every node $e_{\pi}$ is connected to node $A'_d$ if the trip $\pi$ ends in depot $d$. \SP{ The set of arcs connecting depots and trips is denoted $\mathcal{A}_D$}.
We also add extra arcs $(A_d, A'_d)$ with infinite capacity and zero savings, to represent the case where a vehicle is not used and stays in the depot. \SP{ We denote this set of arcs by $\mathcal{A}_{\infty}$; and the set of all arcs by $\mathcal{A} = \mathcal{A}_{\pi} \cup \mathcal{A}_D \cup \mathcal{A}_{\infty}$}. Finally, we draw the nodes in a time-space network, where the x-axis represents the time of day, and the y-axis represents the depots.

\begin{figure}
\centering
\includegraphics[width=.9\textwidth]{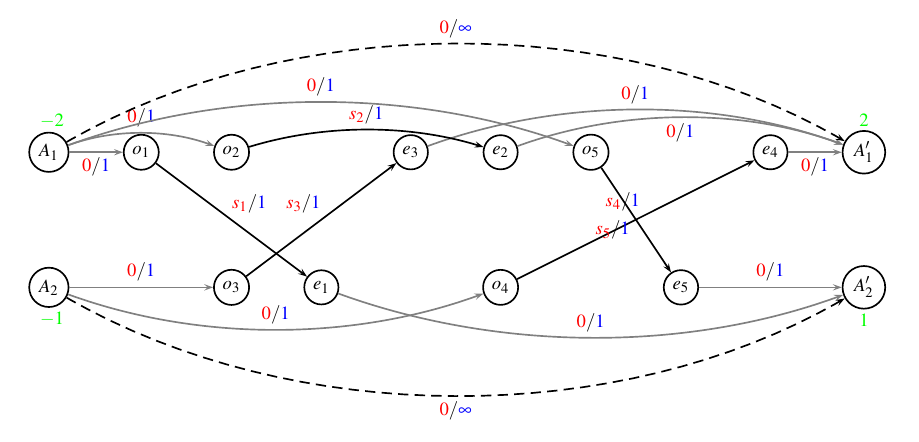}
\caption{The underlying graph of the minimum-cost flow formulation of the vehicle-sharing problem with 
one shared vehicle type, five trips, and two depots. Nodes $A_d,A'_d$ represent depots where the available vehicles are stored at the beginning and end of the time horizon. In our example we have $\delta_1$ = 2 vehicles available at $A_1$ and $\delta_2$ = 1 car at $A_2$, the same amount of vehicles has to be returned in the evening to $A'_1$ and $A'_2$. Nodes $o_{\pi}$ and $e_{\pi}$ give start and end points of a trip $\pi$. Finally, each arc represents a trip $\pi$ with a given saving $s_{\pi}$ and capacity.
The x-axis represents the time of day, and the y-axis represents the depots.} 
\label{fig:minflow}
\end{figure}

Figure \ref{fig:minflow} shows a simple example in which we have two depots, and five trips. We assume that the first depot has two vehicles available in the morning, and two vehicles (not necessarily the same) should be returned to the depot in the evening. Note that we indicate the savings and capacity for each arc in the form (\textit{savings, capacity}). 

Let $V$ be the set of all nodes and let $s_{ij}$ be the savings of a trip going from node $i$ to $j$ (in our auxiliary graph $e_{\pi}$, $o_{\pi}$). Furthermore, let $\delta_i$ be the demand at the depots, being 0 for $e_{\pi}$ and $o_{\pi}$. Parameter $u_{ij}$ gives the capacity of an arc, which is $1$ for all arcs \SP{in $\mathcal{A}_{\pi} \cup \mathcal{A}_D$ and infinite for all arcs in $\mathcal{A}_{\infty}$}. 
Finally, the binary decision variables $x_{ij}$ take on value 1 if connection $(i,j)$ \SP{$\in \mathcal{A}$} is chosen, and 0 otherwise.

With this, we show that the vehicle-sharing problem considering one single type of shared vehicle (\singlesharing) can be modeled as the maximization equivalent to a minimum-cost flow problem, formulated in model~\eqref{eq:minflow-of}-\eqref{eq:minflow-bin}.

\begin{eqnarray}
  \mbox{max} &  \label{eq:minflow-of}
                \dis \sum_{(i,j) \in \SP{\mathcal{A}}} s_{ij} x_{ij} \\
  \mbox{s.t} & \label{eq:minflow-demand}
                \dis \sum_{\SP{j | (j,i) \in \mathcal{A}}} x_{ji} - \sum_{\SP{j | (i,j) \in \mathcal{A}}} x_{ij} = \delta_i 
                & \forall i \in V \\
                \label{eq:minflow-capacity}
             & \dis x_{ij} \leq u_{ij} & 
             \SP{ \forall (i,j) \in \mathcal{A}} \\
             \label{eq:minflow-bin}
             & \dis x_{ij} \geq 0, \SP{\text{integer}} & 
              \SP{ \forall (i,j) \in \mathcal{A}}
              \label{eq:minflow-nonneg}
\end{eqnarray}

The objective function~\eqref{eq:minflow-of} maximizes savings.
Constraints~\eqref{eq:minflow-demand} restrict the out/ingoing vehicles at the beginning/end of the day. Further it assures flow conservation in nodes $i \in V \setminus \{D\}$. 
Constraints~\eqref{eq:minflow-capacity} make sure that at most $u_{ij}$ vehicles cover a certain connection $(i,j)$.

We solve our model as a mixed integer program (MIP), since state-of-the-art solvers are already capable of handling these kinds of problems very efficiently. Nevertheless, we shortly review some of the algorithms that have been widely applied. 
\citet{Fulkerson1962} were first to introduce a combinatorial algorithm for the problem. \citet{Edmonds1972} proposed the scaling algorithm resulting in the first weakly polynomial-time algorithm. \citet{Tardos1985} introduced the minimum cost circulation algorithm which was the first strongly polynomial method.
In the consecutive years many solution approaches evolved. Scaling techniques have shown to be promising \citep{Edmonds1972,Goldberg1990,Goldberg1997,Buennagel1998}. Polynomial in time are also cycle cancelling algorithms \citep{Klein1967,Goldberg1989} or cut cancelling algorithms \citep{Ervolina1993}. Furthermore, the network simplex method was efficiently applied to the maximum flow problem  \citep{Dantzig19633,Kelly1991,Loebel1996} or adaptions of the successive shortest path algorithm \citep{Brunsch2013}.
\citet{Kovacs2015} provides a survey of various algorithms and present an overview of their respective complexity.

\subsection{The vehicle-sharing problem with multiple types of shared vehicles (\multisharing)}
\label{sec:multisharing}

In what follows, we do not only consider one type of shared vehicle but multiple ones. Note that shared vehicles can be different types of cars but also bikes or any other MOT.

We start with the previously described graph. To model the vehicle-sharing problem with multiple types of shared vehicles (\multisharing), we duplicate the sources and sinks, since we have different MOT options and \SP{add} super-nodes where the MOTs start/end. We model a super-source $M^k$ for each $k \in K'$ where the set $K' \SP{\subseteq } K$ gives the set of shared MOTs. In our example, \SP{$K' = \{1,2\}$, where MOT $k = 1$ denotes combustion engine cars and MOT $k = 2$ battery electric cars. For each $k \in K'$ we create a super-source node: $M^1$ for combustion engine cars, 
and $M^2$ for battery electric cars.}

\begin{figure}
\centering
\includegraphics[width=.9\textwidth]{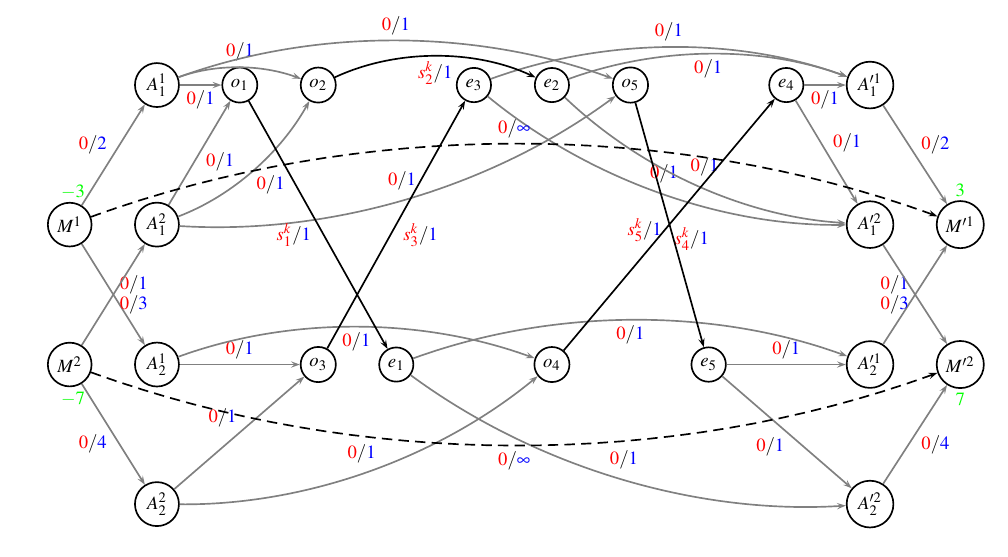}
\caption{The underlying graph of the multi-commodity flow formulation of the vehicle-sharing problem with
two shared types of vehicles, five trips $\pi$, and two depots $d$. Nodes $M^k,M'^k$ represent super-nodes where the available shared vehicles are stored at the beginning and end of the time horizon and then distributed to the respective depot nodes $A^k_d, A'^k_d$. We have 3 vehicles of type $1$  and 7 vehicles of type $2$; 2  \SP{vehicles} and 1 \SP{vehicle} of type $1$ are distributed to depot 1 and 2, respectively; 3 and 4 \SP{vehicles} of type $2$ to depot 1 and 2, respectively. Nodes $o_{\pi}$ and $e_{\pi}$ give start and end points of a trip $\pi$. Finally, each arc gives its respective savings and capacity.
The x-axis represents the time of day, and the y-axis represents the depots.}
\label{fig:multicom}
\end{figure}

In a similar way we add super-sinks $M'^k$. The set of all super-nodes, thus $M^k \cup M'^k$, is denoted as \SP{$\mathcal{M}$. }
We then construct start and end depot nodes $A^k_d$,$A'^k_d$ to where we connect the respective $M^k$ and all origins $o_{\pi}$ and end nodes $e_{\pi}$ of trips $\pi$ \SP{$\in \mathcal{A}_{\pi}$}, respectively. \SP{The set of arcs connecting super-source nodes and super-sink nodes with the depot nodes is denoted by $\mathcal{A}_M$. The set of arcs connecting start and end depot nodes with trips is denoted by $\mathcal{A}_D'$. The set of all arcs is then defined as $\mathcal{A}' = \mathcal{A}_{\pi} \cup \mathcal{A}_D' \cup \mathcal{A}_M$}
Drawing the nodes in a time-space network, Figure \ref{fig:multicom} shows the simple case where we have two shared types of vehicles.

We show that the problem can then be solved as an integer multi-commodity flow problem, where arc savings $s_{ij}^k$ depend on commodity $k$. 
\SP{Each type of shared vehicle is represented by a commodity. In practice, and like in our example, this could be electric vehicles and conventional combustion engine cars, but it could also be bikes or any other MOT of which only a limited number is available.}
\SP{The number available of each $k \in K'$ is denoted by $\Delta_k$. This is equivalent to the demand at the sink nodes in a traditional multi-commodity flow setting. Furthermore, the capacity of each arc $u_{ij}$ defines how many commodities (vehicles) are allowed to traverse it. In our case this is 1 except for those arcs which connect the super-source nodes with the start depots and the end depots with the super-sink nodes. In this case, the capacity defines how many vehicles of MOT $k \in K'$ are available at each of the depots. Finally, } 
let \SP{integer variable} $x^k_{ij}$ \SP{take on the number of vehicles of shared MOT $k \in K$ traversing arc $(i,j)$. Using this notation, we formulate the \multisharing as follows:} 

\begin{eqnarray}
\label{eq:mult-of}
  \mbox{max} & \dis \sum_{k \in K} \sum_{\SP{(i,j) \in \mathcal{A}'}} s_{ij}^k x_{ij}^k \\
  \mbox{s.t} & \dis \sum_{i | \SP{(i,j) \in \mathcal{A}'}} x_{ij}^k -
                    \sum_{i | \SP{(j,i) \in \mathcal{A}'}} x_{ji}^k = 0 &
             \forall  k \in K, j \in V \setminus \{M\} \label{m1} \\
             & \dis \sum_{j | \SP{(i,j) \in \mathcal{A}'}} x_{ij}^k - 
                    \sum_{j | \SP{(j,i) \in \mathcal{A}'}} x_{ji}^k = \Delta_k &
              \forall i \in M^k, k \in K \label{m2} \\
             & \dis \sum_{i | \SP{(i,j) \in \mathcal{A}'}} x_{ij}^k -
                    \sum_{i | \SP{(j,i) \in \mathcal{A}'}} x_{ji}^k = \Delta_k &
             \forall  j \in M'^k, k \in K \label{m3} \\
            & \dis \sum_{k \in K} x_{ij}^k \leq u_{ij} &
              \forall \SP{(i,j) \in \mathcal{A}'} \label{m4} \\
            & \dis x_{ij}^k \geq 0, \SP{\text{integer}} &
              \forall k \in K, \SP{(i,j)\in \mathcal{A}'} \label{m5}
\end{eqnarray}
Objective function~\eqref{eq:mult-of} maximizes the savings.
Equations~\eqref{m1} give the flow conservation constraints for all nodes except the sources and sinks.
Constraints~\eqref{m2} and \eqref{m3} restrict the number of shared MOTs.
Constraints~\eqref{m4} give the capacity restriction on the arcs.
Finally, constraints~\eqref{m5} define the domains of the decision variables.

The formulation above is polynomial in the size of the constraints, having $|K| \cdot |\mathcal{A}'|$ variables 
and $|\mathcal{A}'|+|K| \cdot |V|$ constraints. However, large-scale problems may be challenging to be solved. Therefore, efficient solution algorithm\SP{s} have been applied such as Lagrangian relaxation \citep{Retvari2004, Babonneau2006}, adapted branch-and-bound \citep{Barnhart2000}, Dantzig-Wolfe decomposition \citep{Karakostas2008} and column-generation  \citep{Tomlin1966, Barnhart1994}. 
Nevertheless, we solve the models as a MIP as state-of-the art commercial solvers are able to solve problems of limited size (like in our case) within seconds.

\section{Computational results} \label{sec:compresults}

We provide computational results using the above presented models for the vehicle-sharing problem. The models are implemented in C++ and solved with CPLEX 12.9. Tests are carried out using one core of an Intel Xeon Processor E5-2670 v2 machine with 2.50 GHz running Linux CentOS 6.5. Tests are conducted on a number of generated instances varying in size and complexity. 

In the following, we give a short introduction to the instance set. Afterwards we provide the results of our computational study for \singlesharing\ and \multisharing. We further present results of varying objective functions and restricted sets of MOTs for individual users. Lastly, we comment on the results and give some managerial insights.

\subsection{Test instances}\label{sec:data}

\SP{During the project}, 
benchmark instances based on available demographic, spatial and economic data of the city of Vienna, Austria \SP{were generated}. Five different MOTs are considered: vehicles (\SP{internal} combustion engine vehicles (cars) and \SP{battery} electric vehicles (ecars)), walk, bike, public transportation and taxi. 
For each mode of transport $k \in K$, 
distances, time and cost between all nodes \SP{are computed}.
\SP{For this purpose,} 
the aerial distance between two locations \SP{is taken} which \SP{is} 
then multiplied by a constant sloping factor for each MOT $k$ in order to account for longer/shorter distances of the respective mode.
Moreover, we have emissions per distance unit, average speed, cost per distance and cost per time as well as additional time needed for, e.g., parking a car, for each $k \in K$. The cost of time is a fixed value based on the \SP{median} gross salary \SP{of 2017} \SP{\citep{einkommen2018}} including additional costs for employers in Austria, \SP{which were added to the base value}. The objective function \SP{coefficients} result from these values.
The values of the different parameters are given in the Appendix (Table~\ref{tab:instance-data}).

Each generated instance represents a distinct company operating two offices and consisting of a predefined set of 
employees (or users), $p \in P$. The locations of the offices (depots) are based on statistical data of office locations in Vienna \citep{StadtWien2016} 
placed in the geometric centers of all 250 registration districts, like in  \citet{Knopp2018}. 


Companies are defined by a fixed number of \SP{employees or} users $u \SP{ = |P|}$. Note that one person may have more than one trip assigned. Therefore, the number of users $u$ does not equal the number of trips (\SP{represented by} arcs) in the graph. In Table~\ref{tab:usertrips-data} we provide an overview of the average number of trips per user. On average each user takes about 1.5 trips during the planning horizon \SP{and each trip consists of one or two tasks (or meetings), taking place at different locations}.

\begin{figure}[t]
    \centering
    \includegraphics[width=0.7\textwidth]{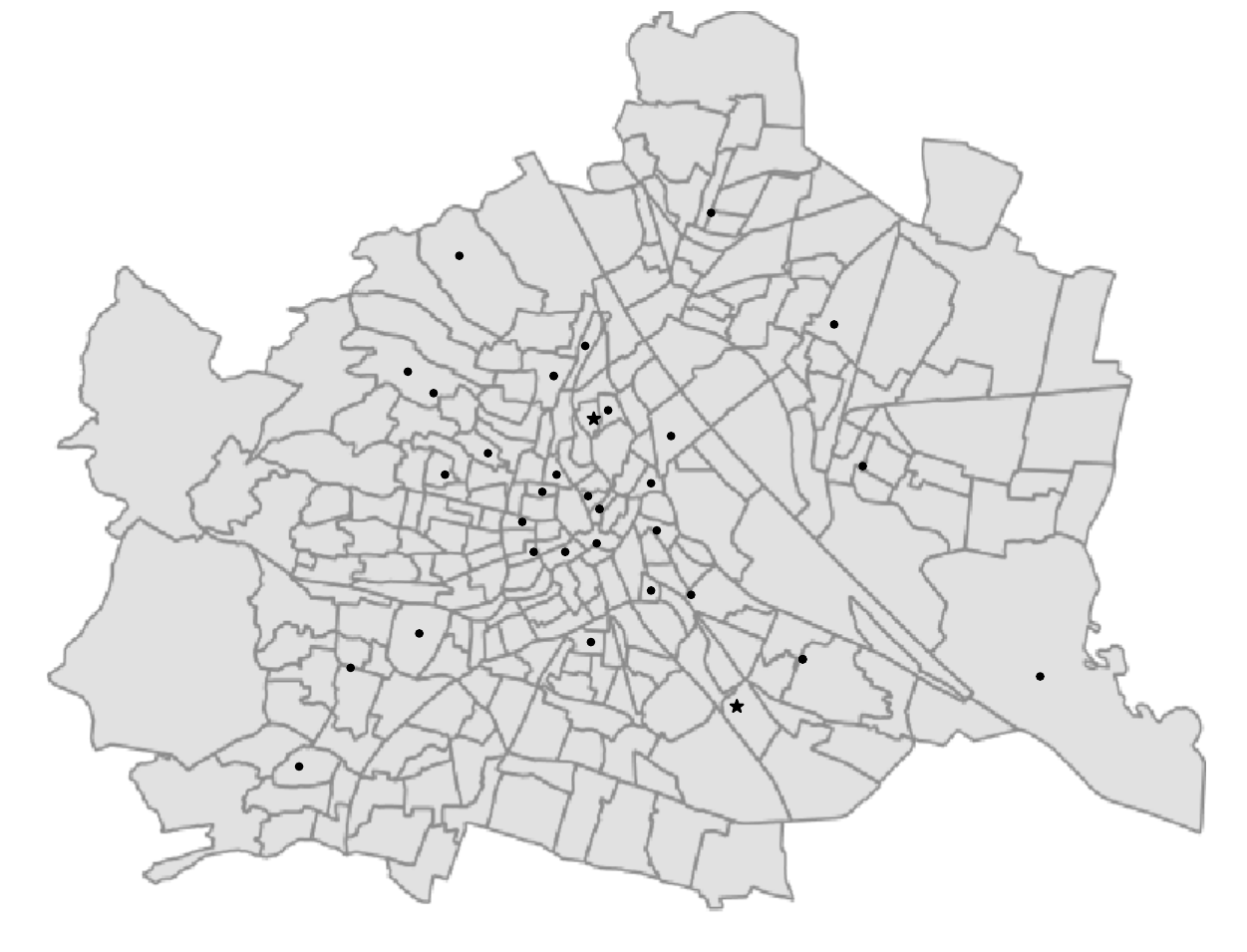}
    \caption{\SP{Instance E\_20\_7, map source: Stadt Wien - \url{https://data.wien.gv.at}, 
(\href{https://creativecommons.org/licenses/by/4.0/deed.de}{CCBY4.0})}}
    \label{fig:E-20-7}
\end{figure}

The number of meetings and their time and location are randomly generated based on historic statistical data \citep{Knopp2018}. 
\SP{E.g., a company with 20 employees as represented by instance E\_20\_7 has two office locations, one in district "Wallensteinstrasse" and one in district "Hasenleiten" and the task locations are distributed across Vienna as shown in Figure \ref{fig:E-20-7}; asterisks indicate office locations, dots are task or meeting locations.}

We define a time horizon of one day where each user has an assigned set of meetings distributed over the day. We calculate savings based on the cheapest other MOT, whereas we always use publicly available MOTs (public transportation, bike, taxi) to be the cheapest other possible alternative. \SP{Using this approach, across all considered instances (we generate 10 instances per instance group), the average distance of those trips where bike is the cheapest alternative, amounts to 13.6 km, the average distance of those trips where public transport is the cheapest alternative is 29.9 km and the average distance of those trips where walking is the cheapest alternative amounts to about 0.5 km.}

\begin{table}[htbp]
  \centering
  \caption{Average number of trips for each instance group with $u$ users.}
    \begin{tabular}{lrrrrrrr}
    \toprule
    \textit{u} & \textit{20} & \textit{50} & \textit{100} & \textit{150} & \textit{200} & \textit{250} & \textit{300} \\
    trips & 31    & 76    & 147   & 218   & 287   & 358   & 427 \\
    trips / $u$ & 1.54  &              1.52  &              1.47  & 1.45  &              1.44  &              1.43  &              1.42  \\
    \bottomrule
    \end{tabular}%
  \label{tab:usertrips-data}%
\end{table}%

A more detailed instance description can be found in \citet{Enzi2020}. \citet{Knopp2018} base their instance generation on the same idea, and provide a detailed description at the end of their paper. Instance sets are made publicly available at \url{https://github.com/dts-ait/seamless}. 


\SP{
\subsection{Vehicle-sharing with a single type of shared vehicle (\singlesharing) and with multiple types of shared vehicles (\multisharing)}

In the following we compare the results obtained for the \singlesharing, represented by model \eqref{eq:minflow-of}-- \eqref{eq:minflow-nonneg}, and those obtained by \multisharing, given in model \eqref{eq:mult-of}--\eqref{m5}. We consider two cases of one type of shared vehicle: in \minflowcar\ these are combustion engine cars (\car), in \minflowecar\ we consider electric cars (\ecar) as our shared resource. Furthermore, we consider the case of a mixed fleet (\mix), which consists of combustion engine as well as electric cars. These results are obtained by solving model \multisharing. In order to analyze the impact of different fleet sizes and compositions in combination with different numbers of users, we compute the average cost per instances size for each combination of $m \in \{4,8,20,40\}$ and $u \in \{20, 50,100,150,200,250,300\}$. Walk, bike, public transportation, and taxi are assumed to have no capacity restriction. 
The considered vehicles are equally spread over the two depots. This means that $m/2$ vehicles of type \car, respectively \ecar, are available at each depot in the homogeneous fleet case (\singlesharing), whereas $m/4$ of each type are available in the mixed case ($m/2$ of type \ecar\ and $m/2$ of type \car, resulting in a fleet of size $m$). In Figure~\ref{fig:costcomp}, we compare the different fleet sizes and fleet compositions in terms of their average total costs per instance class and we observe that for a fixed number of vehicles, \ecar\ results in the lowest costs. With $m = 20$ electric cars (\ecar) and $u = 20$ the cost, in comparison to the base case with $m = 0$, can be reduced by $19\%$. As can be seen from the figure, even in the case where as many vehicles as users are available, for some trips it is cheaper to rely on other MOTs, highlighting the advantage of considering alternative MOTs. In the case of $u = 150$ and $u = 300$ with $m = 40$, the reduction in comparison to the base setting ($m = 0$) amounts to at most $17\%$ and $12\%$, respectively. Detailed results for all combinations of $u$ and $m$ can be found in the Appendix in Tables~\ref{tab:appendix:cost-car-percentage-nopref},~\ref{tab:appendix:cost-ecar-percentage-nopref}, and~\ref{tab:appendix:cost-multcomm-percentage-nopref}. In the Appendix (Table~\ref{tab:time-nopref}) we also give an overview of the solution times for \minflowcar\, \minflowecar, and \multcomm. For an increasing number of users $u$, we observe an increase in the times used to solve the models. However, we always stay below 17 seconds of computation time.}

\SP{
\begin{figure}
    \centering
     \includegraphics[width=\textwidth]{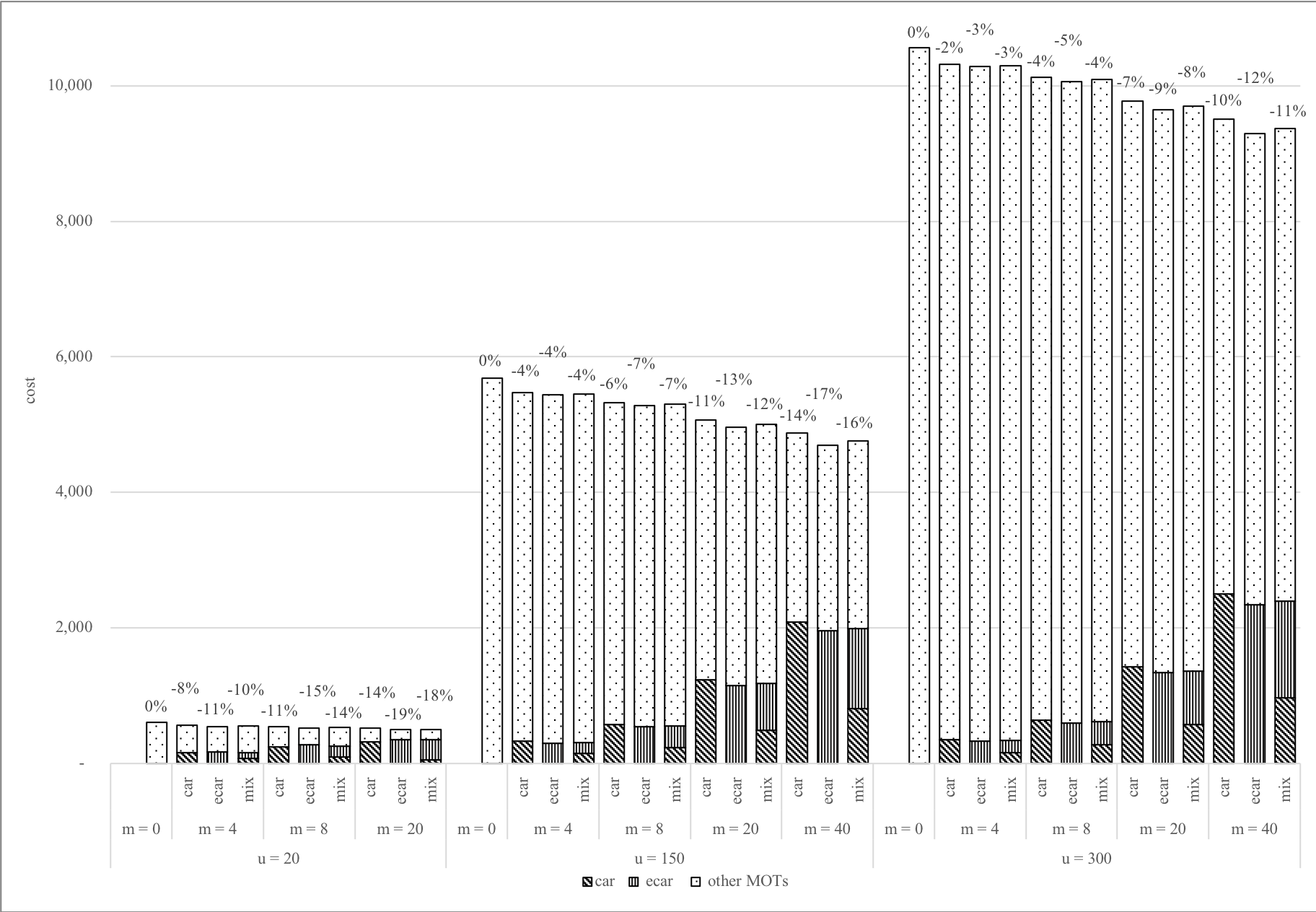}
    \caption{\SP{Total cost breakup into cost of MOTs and cost of vehicles, respectively, for a homogeneous fleet of combustion engine cars (\car), battery electric cars  (\ecar), and a mixed vehicle fleet (\mix) for an increasing number of $u=20,150,300$ and cars $m=4,8,20,40$ in comparison to the base setting ($m = 0$).} }
    \label{fig:costcomp}
\end{figure}
}

\begin{figure}
    \centering
    \begin{subfigure}[c]{0.45\textwidth}
    \includegraphics[width=\textwidth]{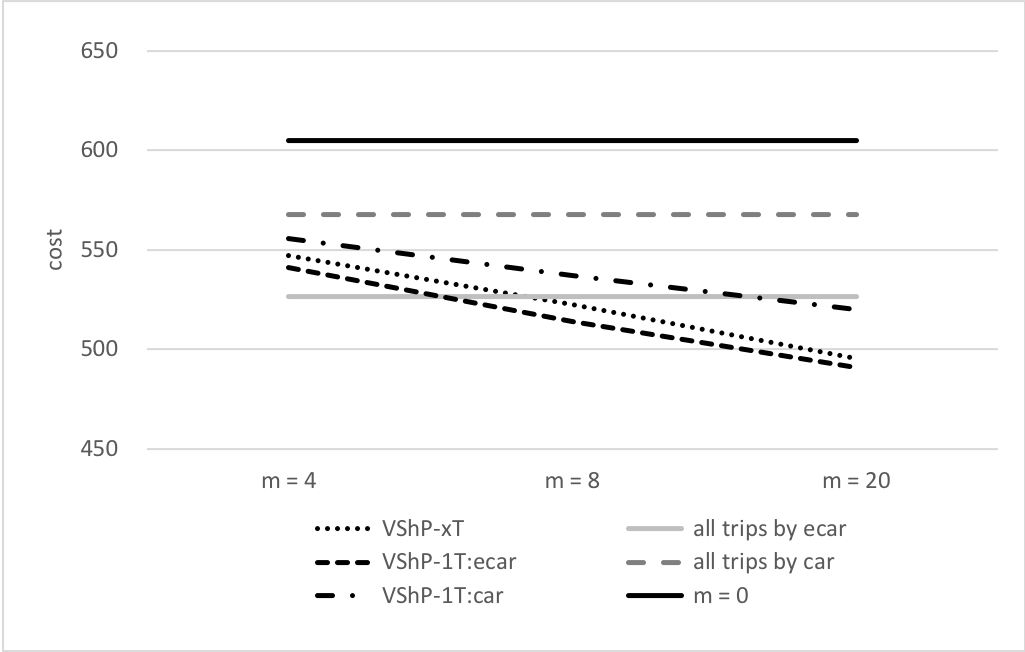}
    \subcaption{u = 20}
    \end{subfigure}
    \begin{subfigure}[c]{0.45\textwidth}
    \includegraphics[width=\textwidth]{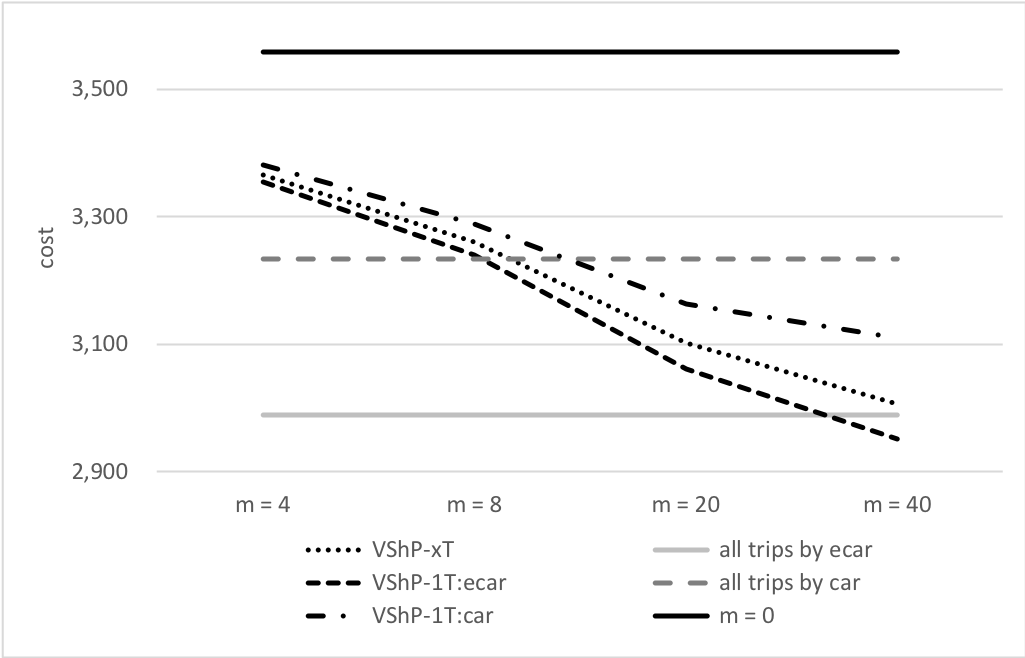}
    \subcaption{u = 100}
    \end{subfigure}
    \begin{subfigure}[c]{0.45\textwidth}
    \includegraphics[width=\textwidth]{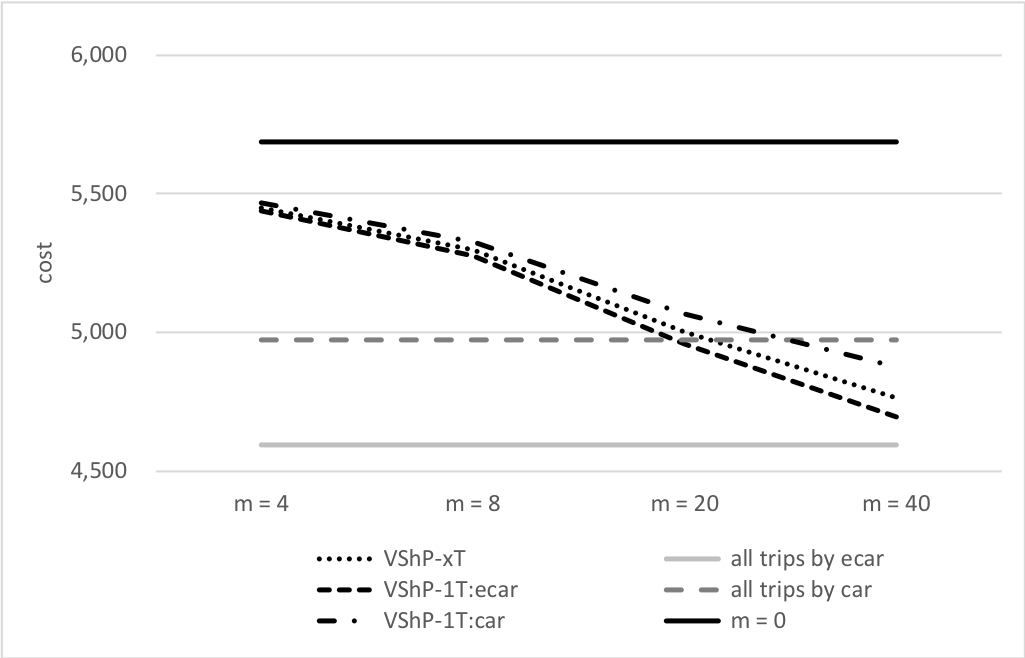}
    \subcaption{\SP{u = 150}}
    \end{subfigure}
    \begin{subfigure}[c]{0.45\textwidth}
    \includegraphics[width=\textwidth]{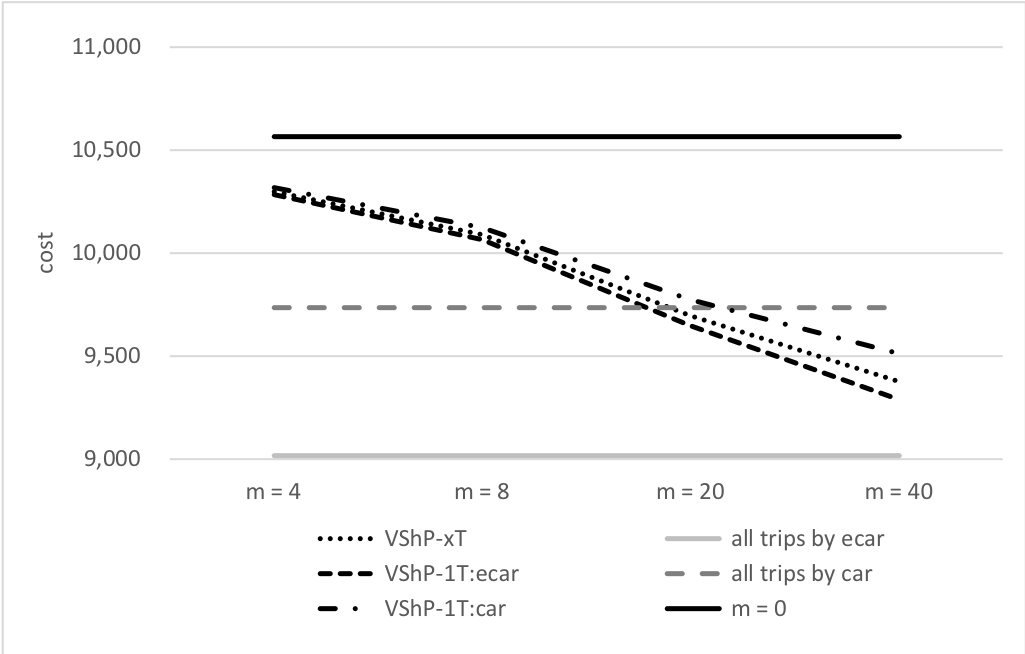}
    \subcaption{u = 300}
    \end{subfigure}
    \caption{Total cost comparison where all trips are either covered by electric cars (\ecar), combustion engine cars (\car) or not by cars at all \SP{($m = 0$)}, and the introduced models (\minflowcar, \minflowecar, \multcomm). The restricted fleet (given on the x-axis, $m=4,8,20,40$) is only applicable to the cases where vehicles are shared (\minflowcar, \minflowecar, \multcomm) as otherwise all trips are covered by the respective MOT.}
    \label{fig:compare-cost}
\end{figure}

\SP{We now compare the different fleet sizes for different users also to the two cases where all trips are covered by electric vehicles and  to the case where all trips are covered by internal combustion engine vehicles.} Figure~\ref{fig:compare-cost} shows the cost of the different cases for $u=20, 100, 300$ \SP{employees} and increasing \SP{fleet size} $m$. 

The respective lines give the cost of the following cases: no trip is covered by a vehicle ($m = 0$), every trip is covered by a combustion engine vehicle (\car), all trips are covered by electric vehicles (\ecar), \minflowcar, \minflowecar\ and \multcomm. Note that the fleet restrictions only apply for \minflowcar, \minflowecar\ and \multcomm.
We observe that it is always most expensive if no trip is covered by a car. In all three figures, the line representing cost of using only \SP{electric vehicles} lies below the line showing cost when using combustion engine vehicles, only. 
When considering $u=20$, \minflowcar, \minflowecar, and \multcomm\ are always cheaper than employing conventional cars only \SP{and result in lower costs than assigning individual electric cars to each user with a shared fleet of only 8 vehicles.} 
\SP{In the case of a larger number of users $u = 100$  and $u = 300$, a shared fleet in combination with other MOTs is less expensive than individually assigned combustion engine cars, if a vehicle fleet of $m = 20$ electric vehicles is deployed. For  medium-sized companies with $u = 100$ employees, a shared electric fleet of only $m = 40$ vehicles in combination with the considered other MOTs, results in lower average costs than individually assigned electric vehicles. Summarizing these findings, from a company perspective, a shared electric vehicle fleet (\minflowecar) having a size of about 40\% of the total number of users in combination with bike, walking, public transport, and taxi results in lower or equal costs than covering all trips by electric vehicles.}

\SP{When comparing the costs of the three settings (\minflowcar, \multcomm, and \minflowecar) on an aggregated level (across all fleet size settings),} \minflowecar\ is the cheapest, as we have already seen in the previously discussed figure. \multcomm\ shows on average slightly higher cost. Lastly, as expected, the case where only combustion engine cars are employed in a pool of shared vehicles, is the most expensive alternative, ranging up to 1.05 times the cost of \minflowecar. \SP{Table~\ref{tab:comparison-models} in the Appendix compares the cost of the three models \minflowcar, \multcomm, and \minflowecar, for all considered user settings, using \minflowecar\  as the base.} 

\subsection{\SP{Vehicle usage}}
\begin{figure}
    \centering
     \includegraphics[width=0.8\textwidth]{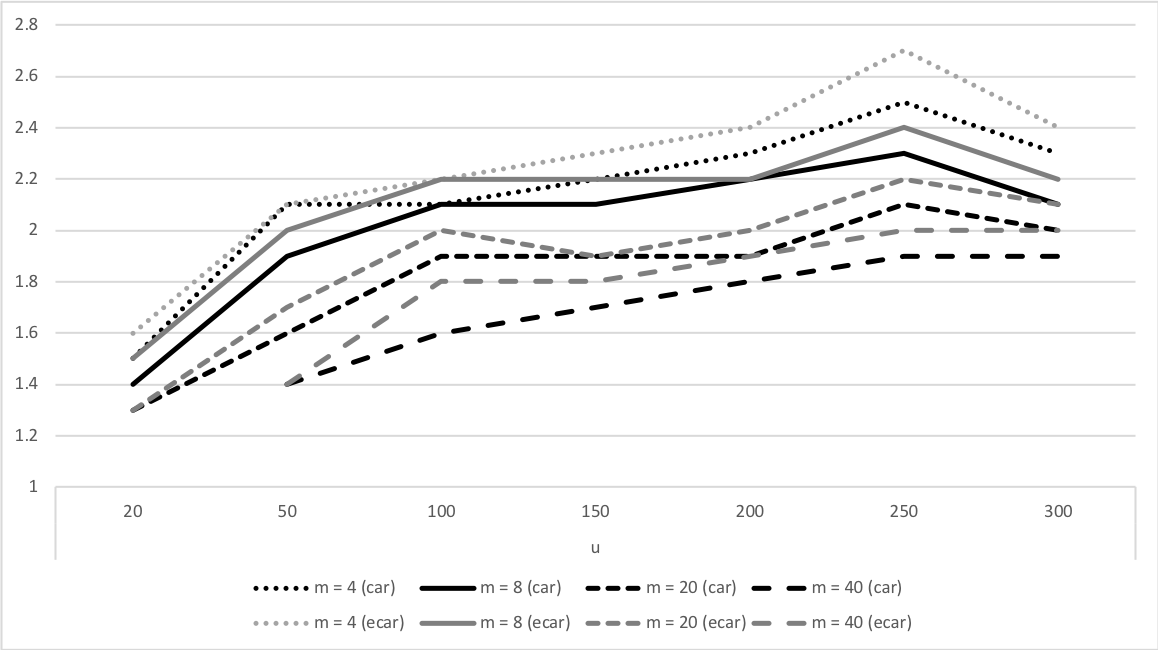}
    \caption{\SP{Comparison of the average number of trips per vehicle for an increasing number of users $u$ and vehicles $m$ for a single type of shared vehicle,  solving \minflowcar\ or \minflowecar.
    \label{fig:tripsPercar}
    }}
\end{figure}

\SP{In order to evaluate the usage of the vehicle fleet, we compute the average number of trips covered by car.} 
\SP{Figure~\ref{fig:tripsPercar} shows the average number of trips for a homogeneous shared fleet, with either conventional combustion engine vehicles (\minflowcar, car) or battery electric vehicles (\minflowecar, ecar) for an increasing number of vehicles $m$ and users $u$. } 
We observe that with an increasing number of users $u$ the average number of trips for a car is also rising. This is because the model aims to cover as many trips by car as possible. With an increasing number of users but the same number of vehicles in the system, 
more trips \SP{are situated on average} on one of the few cars available. \SP{Furthermore, } the average number of trips is higher when fewer cars are available. \SP{Since not all trips are compatible in terms of their timing, duration, and depot location, the more vehicles are available for the same number of users, the fewer beneficial combinations remain and the fewer trips per vehicle are performed.} We observe this for both variants, \minflowcar\ and \minflowecar. However, overall \minflowecar\ shows a higher average \SP{number of} trips per car.

Table~\ref{tab:multcomm-trips} shows the average number of trips a car is taking for an increasing number of users $u$ and cars $m$ for \multcomm. The results are split into values for the different car types. We can see that the average number of trips per electric car (\ecar) is always greater than the number of trips for the other type. This means, that if possible, 
more trips are assigned to electric vehicles. 
Moreover, we can again observe an increase in the average number of trips per vehicle for a higher number of users $u$ as well as \SP{a} smaller number of vehicles $m$.

\begin{table}[htbp]
  \centering
  \caption{Average number of trips per vehicle for an increasing number of users $u$ and vehicles $m$ for \multcomm. \SP{Note that since we consider two types of vehicles, there are $m/2$ of each type available in the system.}
  }
    \begin{tabular}{rlcccccccccc}
    \toprule
 \multicolumn{1}{l}{\textit{m}}  & type & \multicolumn{1}{l}{\textit{u=}}  &   \multicolumn{1}{r}{20} & \multicolumn{1}{r}{50} & \multicolumn{1}{r}{100} & \multicolumn{1}{r}{150} & \multicolumn{1}{r}{200} & \multicolumn{1}{r}{250} & \multicolumn{1}{r}{\textit{300}} \\
\midrule
4 &  \car & &                1.5  &                1.7  &                1.7  &                1.9  &                2.1  &                2.1  &                1.9  \\
 &      \ecar  &      &           1.8  &                2.5  &                2.6  &                2.6  &                2.6  &                3.0  &                2.8  \\
 8 &  \car & &                1.3  &                1.6  &                1.7  &                1.6  &                1.8  &                1.8  &                1.7  \\
 &       \ecar  &    &            1.6  &                2.3  &                2.6  &                2.6  &                2.7  &                2.8  &                2.6  \\
 20 &  \car & &                0.4  &                1.2  &                1.6  &                1.4  &                1.4  &                1.6  &                1.5  \\
 &        \ecar  &  &              1.5  &                2.1  &                2.4  &                2.4  &                2.5  &                2.7  &                2.5  \\
 40 &  \car & &                 &                0.7  &                1.3  &                1.3  &                1.4  &                1.4  &                1.4  \\
 &      \ecar  &     &            &                1.7  &                2.1  &                2.1  &                2.3  &                2.5  &                2.4  \\
    \bottomrule
    \end{tabular}%
  \label{tab:multcomm-trips}%
\end{table}%

\subsection{Incorporating user preferences}

\SP{Assuming that for a successful introduction of a shared mobility system, the consideration of user preferences with respect to the different MOTs is detrimental, we introduce user preference information into our models as follows:} every user $p$ \SP{is associated} with a set $K^p \subseteq K$ of possible modes of transport that can be used, reflecting her preferences. Depending on the user that is covering a trip $\pi$, we can then define a set of modes of transport possible to be assigned to a trip $K^{\pi} \subseteq K$. Note that if a MOT is not in the respective set $K^{\pi}$ we impose a penalty and set $C_{\pi}^k = \infty$. 
We define seven different cases aiming to represent differences in the preference distribution.

For the first case, prefVar0, we make use of available statistical data representing the working population of Vienna. For this group, we define different combinations of possible accepted MOTs in the instance generation: generic, motorised only, no public transportation, no motorised, cars only, public transportation only and bike only. For each of them we have a probability for female and male users, where we have [0.19, 0.03, 0.01, 0.04, 0.18, 0.42, 0.13] and [0.18, 0.03, 0.02, 0.03, 0.26, 0.35, 0.13], respectively \citep{stadt_wien}. We assume that 53\% of the working population is male, and 47\% female \citep{statstik_austria_2017}. Further, we incorporate the probability that 87\% of them have a driving license and 13\% are not allowed to drive a car \citep{bmvit}. 
The combinations are then chosen randomly based on the set probability distribution. We assume that if a user includes a combustion engine car in her set of MOTs, then she will also have the electric car and vice versa.
For the other cases, prefVar1-prefVar6, we adopt more straightforward strategies to represent the preferences of the users. Depending on the variant, we define a fixed percentage of users with a given setting. We say this may either be mixed (= accepting all MOTs), cars only or other MOTs except cars (= no cars). Let us assume an instance with 20 users and 40\% mixed, 40\% cars only and 20\% other MOTs only. Then users 1-8 accept all MOTs, users 9-16 only cars and users 17-20 anything but cars. Table~\ref{tab:prefVars-data} shows the setting of each of the applied variants. \SP{We note that the considered preference settings are by no means exhaustive but they allow us to get some insights into the impact of considering user preferences on the total costs from a company perspective.}

\begin{table}[]
    \centering
    \caption{Categorization of the different preference variants. Percentage of the users with the respective set of accepted MOTs, where  (1) all: no restricted set is applied, user takes all MOTs, (2) cars only: the user only wants to drive by car, (3) no cars: no cars are given in the restricted set, only other MOTs are accepted.}
    \begin{tabular}{rccc}
    \toprule
    variant     &   all   &   cars only   &   no cars  \\
    \midrule
    prefVar0 & \multicolumn{3}{c}{see text}    \\
    prefVar1     &  40\%    &   40\%        &   20\%     \\
    prefVar2     &  10\%    &   10\%        &   80\% \\
    prefVar3     &  25\%    &   25\%        &   50\% \\
    prefVar4     &  0\%    &   80\%        &   20\% \\
    prefVar5     &  0\%    &   20\%        &   80\% \\
    prefVar6     &  0\%    &   50\%        &   50\% \\
    \bottomrule
    \end{tabular}
    \label{tab:prefVars-data}
\end{table}{}

\begin{figure}
    \centering
    \begin{subfigure}[c]{0.48\textwidth}
    \includegraphics[width=\textwidth]{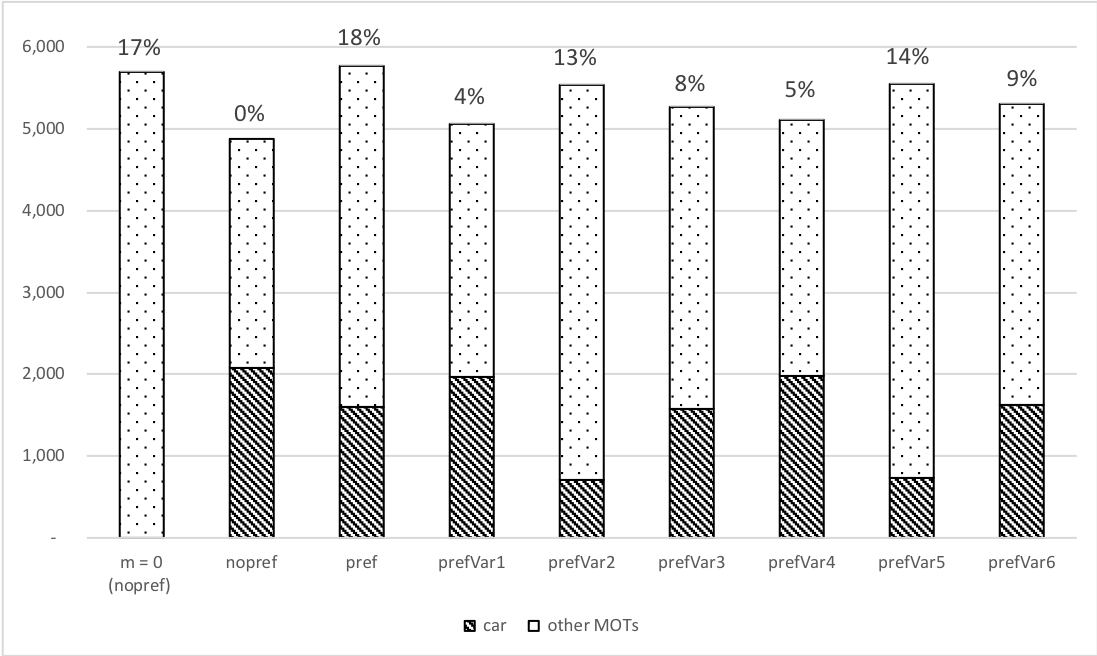}
    \subcaption{u = 150, \minflowcar}
    \end{subfigure}
        \begin{subfigure}[c]{0.48\textwidth}
    \includegraphics[width=\textwidth]{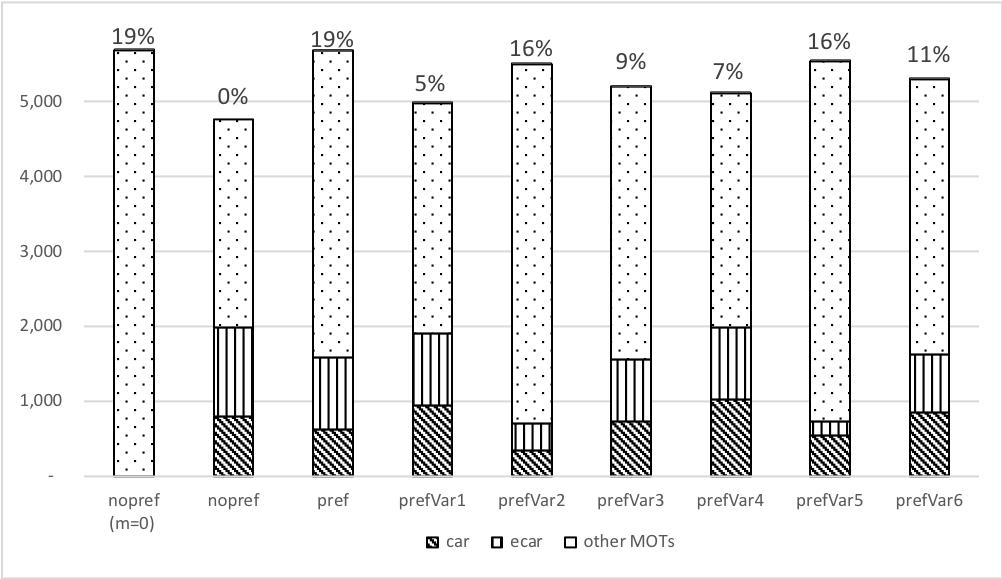}
    \subcaption{u = 150, \multcomm}
    \end{subfigure}
    \caption{\SP{Total cost breakup into cost of MOTs and cost of vehicles (\car\ and \ecar) for $u=150$ and $m=40$ for different variants of MOT-preference settings.}}
    \label{fig:pref-u150}
\end{figure}

\SP{We now analyze the impact of the different preference settings in more detail for two instance classes, $u = 150$ and $u = 300$ both for $m = 40$.
In a first step, we analyze instance class $u = 150$ and we compare the different preference settings to the base setting with a homogeneous vehicle fleet (\minflowcar) and with a heterogeneous vehicle fleet (\multcomm). Figure~\ref{fig:pref-u150} provides the respective results. The percentages above the bars indicate the cost increase with respect to the base setting without preferences. We observe that for instance class $u =  150$ with $m = 40$ vehicles, incorporating preferences increases the costs to about the $m = 0$ setting where no trip is covered by a car. This setting is $17\%$ more expensive than the $m = 40$ setting with a single type of shared vehicle (car) and $19\%$ more expensive than the $m = 40$ setting with a mixed fleet of conventional and battery electric vehicles. Preference setting prefVar1 results in the smallest increase in total costs followed by prefVar4. This result is not surprising since in both cases only 20\% of the users do not accept to use a car. However, in setting prefVar4, there is less flexibility, since 80\% of the users are cars only users, which enforces the use of cars even for trips where this is potentially not efficient.}

\begin{figure}
    \centering
    \begin{subfigure}[c]{0.48\textwidth}
    \includegraphics[width=\textwidth]{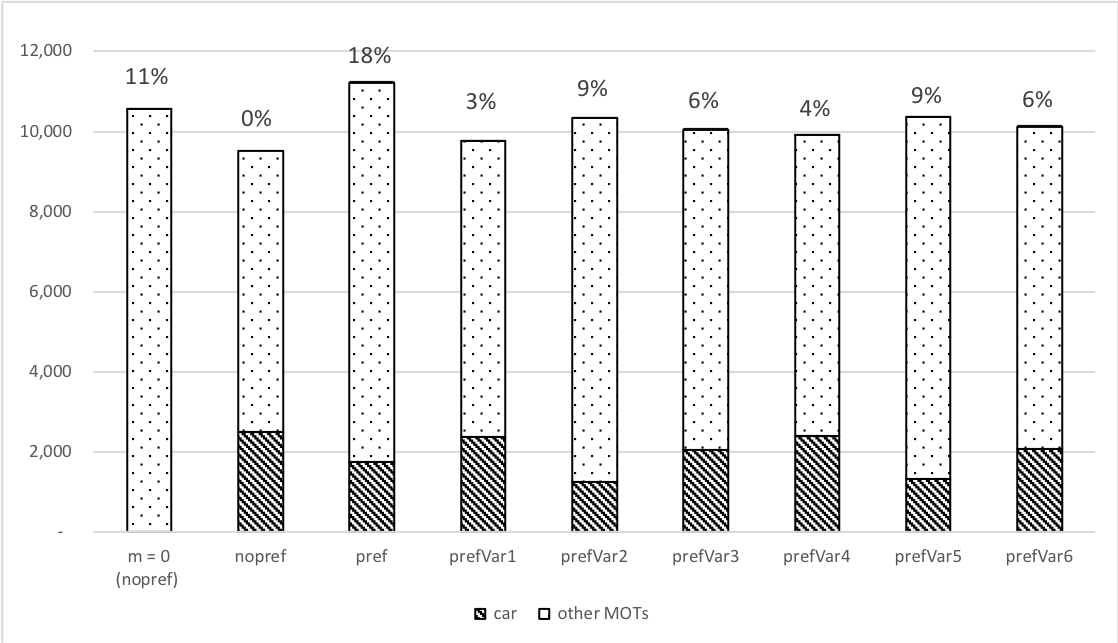}
    \subcaption{u = 300, \minflowcar}
    \end{subfigure}
        \begin{subfigure}[c]{0.48\textwidth}
    \includegraphics[width=\textwidth]{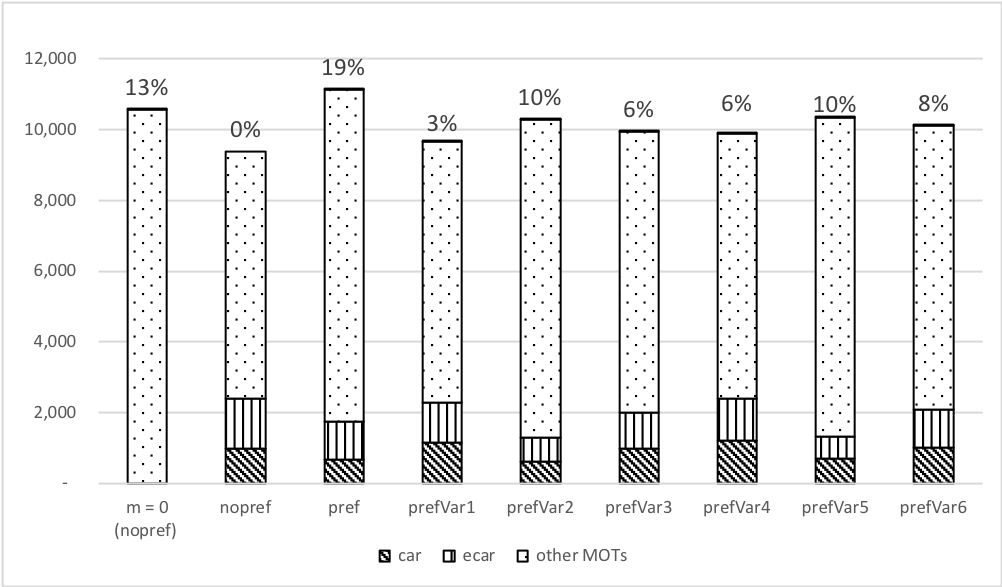}
    \subcaption{u = 300, \multcomm}
    \end{subfigure}
    \caption{\SP{Total cost breakup into cost of MOTs and cost of vehicles (\car\ and \ecar) for $u=300$ and $m=40$ for different variants of MOT-preference settings.}}
    \label{fig:pref-u300}
\end{figure}

\SP{To further analyze if the above observations also hold for larger instances, we also plot the same information for instance class $u = 300$ with $m = 40$ vehicles in Figure~\ref{fig:pref-u300}.
}

\SP{To analyze the changes in cost in further detail,} 
for each variant (prefVar0-prefVar6) we show in Table~\ref{tab:prefVar-car} the average cost of using conventional cars (\car), cost used for all other MOTs and in total for $u=300$ and a homogeneous fleet of $m=40$. \SP{Columns} 
''comp.'' \SP{provide the cost ratio between the respective preference setting and the 
base setting calculated as} (cost of the variant / cost of \minflowcar). We observe that our base setting is the most expensive regarding car usage. In prefVar2, where most of the users prefer all MOTs except cars, we only use 0.59 times the cost of cars compared to the \minflowcar\ setting. Conversely, regarding other MOTs, the simple \minflowcar\ is the cheapest variant \SP{and} prevVar0 hast 1.39 times \SP{higher} cost on average. This comparably big difference in cost is mainly attributable to the more subtle differentiation of the preference settings. As in prefVar0 we also distinguish whether a person would, e.g., only take public transportation. In total we confirm the picture from above, that \minflowcar\ without any restriction, is the cheapest setting, however prefVar1 or prefVar3 only have 1.03 times the cost. Further results can be found in Table~\ref{tab:av-prefVar-car} in the Appendix.

\begin{table}[htbp]
  \caption{Total cost comparison split into cost of \car (=combustion engine cars) and other MOTs for $u=300$ and $m=40$ for different variants of MOT-preference settings and \minflowcar. Column 'cost' gives the absolute cost of the respective variant, 'comp.' compares the cost to \minflowcar\ where it is set as (cost of the variant / cost of \minflowcar).}
    \footnotesize
    \setlength{\tabcolsep}{2pt}
    \begin{tabular}{lcrrrrrrrrrrrrrr}
    \toprule
          & \multicolumn{1}{c}{\minflowcar} & \multicolumn{2}{c}{prefVar0} & \multicolumn{2}{c}{prefVar1} & \multicolumn{2}{c}{prefVar2} & \multicolumn{2}{c}{prefVar3} & \multicolumn{2}{c}{prefVar4} & \multicolumn{2}{c}{prefVar5} & \multicolumn{2}{c}{prefVar6} \\
          & \multicolumn{1}{l}{cost} & \multicolumn{1}{l}{cost} & \multicolumn{1}{l}{comp.} & \multicolumn{1}{l}{cost} & \multicolumn{1}{l}{comp.} & \multicolumn{1}{l}{cost} & \multicolumn{1}{l}{comp.} & \multicolumn{1}{l}{cost} & \multicolumn{1}{l}{comp.} & \multicolumn{1}{l}{cost} & \multicolumn{1}{l}{comp.} & \multicolumn{1}{l}{cost} & \multicolumn{1}{l}{comp.} & \multicolumn{1}{l}{cost} & \multicolumn{1}{l}{comp.} \\
   \midrule
     \car  &               1,227  &                   872  &                  0.71  &               1,169  &                  0.95  &                   722  &                  0.59  &               1,030  &                  0.84  &               1,188  &                  0.97  &                   747  &                  0.61  &               1,045  &                  0.85  \\
     other MOTs  &               8,704  &            12,093  &                  1.39  &               9,014  &                  1.04  &               9,669  &                  1.11  &               9,231  &                  1.06  &               9,220  &                  1.06  &               9,682  &                  1.11  &               9,337  &                  1.07  \\
     total  &               9,932  &            12,964  &                  1.31  &            10,184  &                  1.03  &            10,391  &                  1.05  &            10,261  &                  1.03  &            10,408  &                  1.05  &            10,429  &                  1.05  &            10,382  &                  1.05  \\
    \bottomrule
    \end{tabular}%
  \label{tab:prefVar-car}%
\end{table}%

\subsection{\SP{Alternative objective functions: minimizing time instead of cost}}

\SP{As indicated in the introduction, there potentially exists a trade-off between slow but cheap and fast but expensive MOT. In order to understand if such a trade-off can be observed, we now analyze the impact of the time component in further depth by separating it from the other cost components. In the following, we compare results for our combined cost objective function (OF: base) and an alternative objective function, where we only minimize the total time (OF: time). Since OF:time does not distinguish between electric vehicles and combustion engine vehicles, this analysis is based on a homogeneous fleet setting only and, in order to make our solutions comparable, we take the OF:time solutions and compute their costs using OF:base.}

\begin{figure}
    \centering
    \begin{subfigure}[c]{0.48\textwidth}
    \includegraphics[width=\textwidth]{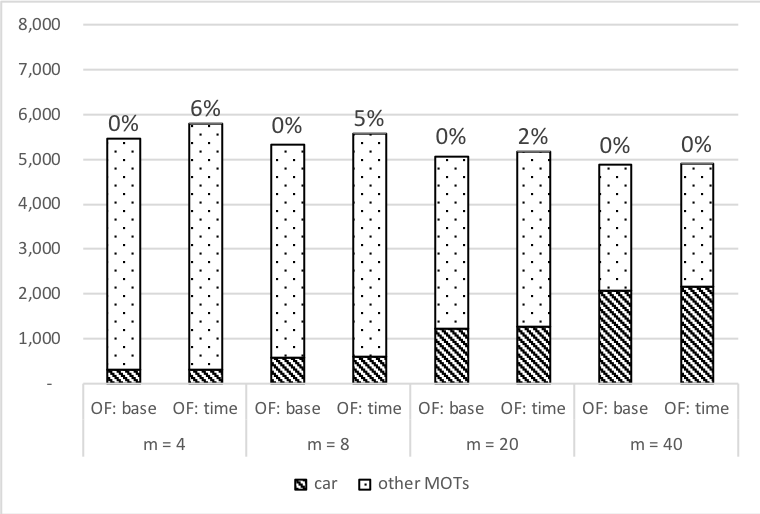}
    \subcaption{u = 150}
    \end{subfigure}
        \begin{subfigure}[c]{0.48\textwidth}
    \includegraphics[width=\textwidth]{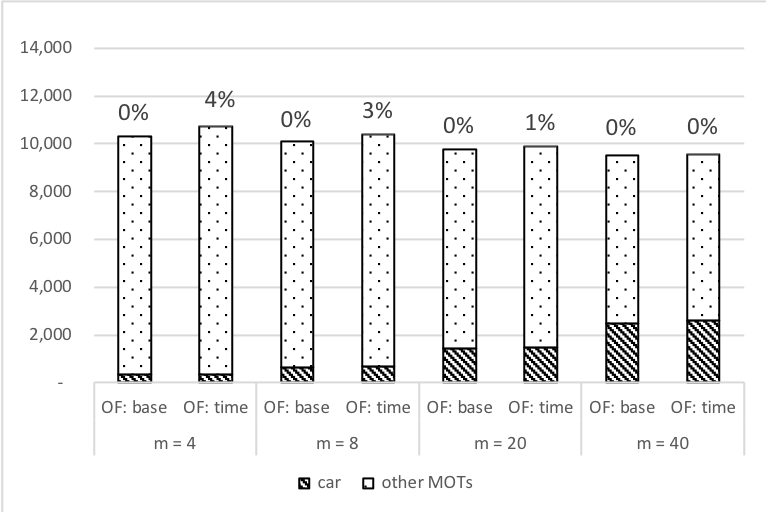}
    \subcaption{u = 300}
    \end{subfigure}
    \caption{Total cost split into cost of MOTs and cost of combustion engine vehicle (\car) for $u = 150, 300$ and different value of $m$ and different objective functions for \minflowcar. OF:base shows the result with the previously introduced objective function, OF:time only considers the time part.
    Note that we solve the models with the different objective functions, but afterwards calculate the total cost to make them comparable.}
    \label{fig:time-car}
\end{figure}

\SP{
Figure~\ref{fig:time-car} shows the composition of the total cost for \minflowcar\ for instances classes $u=150$ and $u = 300$ and different numbers of vehicles $m$. We show the cost share of cars and other MOTs and we observe that the total cost is only slightly higher for OF:time than for OF:cost, which is intuitively right, if time is the major cost driver. On the other hand, this also indicates that the trade-off between slow but cheap and fast but expensive MOTs is not very strong.}

\begin{table}[t]
  \centering
  \caption{Total cost for OF:base and OF:time and their comparison calculated as (OF:time/OF:base), split into cost of MOTs and cost of \car (=combustion engine cars) for an increasing $u$ and averaged over all $m$. OF:base shows the result for the previously introduced objective function, OF:time only considers the time part.
  Note we solve the models with the different objective functions, but afterwards calculate the total cost to make them comparable.}
  \small
    \begin{tabular}{clrrrrrrr}
    \toprule
          & \textit{ u } &                              20  &                              50  &                           100  &                           150  &                   200  &                           250  &                           300  \\
    \midrule
    \multirow{3}[2]{*}{\begin{sideways} \tiny OF:time \end{sideways}} &  \car  &                           333  &                           747  &                       1,004  &                       1,089  &               1,185  &                       1,283  &                       1,277  \\
          &  other MOTs  &                           208  &                           873  &                       2,267  &                       4,270  &               6,100  &                       8,130  &                       8,871  \\
          &  total  &                           541  &                       1,620  &                       3,271  &                       5,359  &               7,285  &                       9,413  &                    10,147  \\
    \midrule
    \multirow{3}[2]{*}{\begin{sideways} \tiny OF:base \end{sideways}} &  \car  &                           258  &                           656  &                           955  &                       1,052  &               1,148  &                       1,255  &                       1,227  \\
          &  other MOTs  &                           275  &                           942  &                       2,281  &                       5,917  &               4,132  &                       7,767  &                       8,704  \\
          &  total  &                           533  &                       1,599  &                       3,235  &                       5,184  &               7,065  &                       9,022  &                       9,932  \\
    \midrule
    \multirow{3}[2]{*}{\begin{sideways} \tiny OF:time / OF:base \end{sideways}} &  \car  &                          1.29  &                          1.14  &                          1.05  &                          1.04  &                  1.03  &                          1.02  &                          1.04  \\
          &  other MOTs  &                          0.76  &                          0.93  &                          0.99  &                          1.03  &                  1.03  &                          1.05  &                          1.02  \\
          &  total  &                          1.02  &                          1.01  &                          1.01  &                          1.03  &                  1.03  &                          1.04  &                          1.02  \\
         &    &      &     &   &   &   &    &   \\  
    \bottomrule
    \end{tabular}%
  \label{tab:time-car}%
\end{table}%

In Table~\ref{tab:time-car} we supplement the above figures with numbers. The table is decomposed into results for OF:base, OF:time and the comparison of the two, where we assume (OF:time/OF:base). The first two are given in absolute numbers, the latter as a ratio of the two. Each partition gives the results of combustion engine cars (\car), other MOTs and in total. The numbers are given on average over all instances and all \SP{fleet} sizes $m$. 
We can see, that using time only as an objective function gives slightly higher overall cost. The smallest difference can be observed for $u=50$ and $u =100$. \SP{It ranges} between 1.01-1.04 times the cost for all instances. 
This difference is mainly driven by 
higher cost of cars with OF:time. 
\SP{Further and more detailed results 
can be found in Table~\ref{tab:timeOF-car} in the Appendix.}

Finally, we compare the average number of trips per car in Table~\ref{tab:time-trips-minflow-av}. 
We observe an increase in trips per car by $10\%$ on average in each instance class ($u$). This indicates that, under the assumption that cars are the fastest MOT, they are chosen slightly more often on average when using OF:time than with OF:base. 


\begin{table}[t]
  \centering
  \caption{Average number of trips per car when solving OF:base and OF:time and their comparison stated as OF:time / OF:base for an increasing $u$ and averaged over all $m$ for \minflowcar.}
    \begin{tabular}{lrrrrrrr}
    \toprule
    \textit{ u = } &                 20  &                 50  &               100  &               150  &               200  &               250  &               300  \\
    \midrule
    OF:time &                1.6  &                1.9  &                2.1  &                2.2  &                2.3  &                2.4  &                2.3  \\
    OF:base &                1.4  &                1.8  &                2.0  &                2.0  &                2.0  &                2.2  &                2.1  \\
    OF:time / OF:base &                1.1  &                1.1  &                1.1  &                1.1  &                1.1  &                1.1  &                1.1  \\
    \bottomrule
    \end{tabular}%
  \label{tab:time-trips-minflow-av}%
\end{table}%

\subsection{Managerial implications and discussion}

We \SP{observe} in all our results, that with a \SP{larger vehicle fleet $m$ } 
(combustion engine vehicles or electric cars), we \SP{obtain solutions of} lower total cost, 
even though vehicles are the most expensive MOT from an operational cost point of view. However, they are, in many cases, the fastest MOTs \SP{and since the time spent traveling is working time and as such has to be paid by the company, the duration of a trip has a strong impact on the total costs. We confirm this result through the comparison of solutions optimizing the cost-based and a time-oriented objective function.} 

Also, whenever possible, electric vehicles are preferred as they have even lower cost but the same speed as conventional vehicles. 
The considered mixed fleet \SP{setting}, $m/2$ electric vehicles and $m/2$ combustion engine vehicles is, as expected, slightly more expensive than \SP{a setting where} only $m$ electric vehicles are employed. \SP{The mixed fleet setting mimics the current situation at many companies, that still operate  a fleet of conventional vehicles which is only gradually replaced by electric cars. 
Operating a shared pool of only combustion engine vehicles is the most expensive option  of the discussed sharing concepts and the least environmental friendly, and thus not recommendable.}

Employing no cars at all, is most expensive. If one decides to go with any (of the presented) sharing concepts, \SP{a vehicle fleet of } 
at least 20\%-25\% of the number of users \SP{is advisable}. E.g., for 20 users this would be 4-5 cars. From there, \SP{in comparison to covering all trips with conventional vehicles,} it starts to be cost efficient to have shared vehicles, and \SP{to} additionally cover trips with other MOTs such as public transportation or bike. 

\SP{Our analysis also shows that the trade-off between fast but expensive and slow but cheap MOTs is quite small, meaning that fast but expensive MOTs are very often the preferred option.}
The shortcuts that can be taken by other MOTs only sometimes outperform the benefits of fast cars. As our instances 
are based on \SP{data from Vienna} 
this makes sense. For longer trips, \SP{outside of metropolitan areas, we expect to observe }
different trade-offs. 

\SP{Concerning vehicle usage, 
we saw in our results, that with the same number of users but a smaller fleet, the trips per vehicle are rising above the average number of trips a user is taking. Furthermore, the average number of trips per vehicle is higher for electric vehicles, due to their lower costs.

Finally, we introduced a set of restricted MOTs based on individual user preferences. As expected, the case where all MOTs are always available for all trips, and thus for all users, is the one with the least cost as it is the least restricted case.} However, also for some of the considered preference settings, our results only show a modest increase in the average costs. 
\SP{The maximum cost increase amounted to $19\%$. 
From a managerial point of view these are the costs for a potentially higher acceptance of a shared system.}

\section{Conclusion} \label{sec:concl}

Inspired by the change of mobility and vehicle-sharing systems we proposed two modeling approaches for a vehicle-sharing problem. In our problem we assume a set of users that have to cover certain trips on a fixed time schedule. These trips are then covered by a certain mode of transport. We assume a restricted available set of shared vehicles, e.g. a pool of cars, which the users may use. Other modes of transport are incorporated without any capacity limits. We aim to assign the restricted resources in the best possible way such that savings (using e.g., a car instead of any other mobility type) are maximized. Note that our initial framework considers a sharing system within a company, however the models can be applied to any community with a closed group of users.

We used two well-known formulations from the literature, namely the maximization equivalent of the minimum-cost flow problem and the multi-commodity flow problem. If we assume only one shared MOT, e.g. cars, we base our formulation on the minimum-cost flow problem. We extend the problem by introducing another type of shared vehicle, and we formulate it as a multi-commodity flow problem where the commodities are the shared vehicles. Note that a shared resource may also be a bike or another MOT.

We further provide managerial insights \SP{considering} combustion engine vehicles and electric cars as our shared vehicles. We show that a shared fleet of electric vehicles contributes most to our objective function. 

Besides this analysis, \SP{our results show the computational advantage of modeling the considered problem settings as minimum-cost flow and  multi-commodity flow problems.} 
Instances with up to 300 users are solved in less than 20 seconds of computing time \SP{with off-the shelf solvers, showing} that our models can be used on a daily operational basis.
As these models are well studied in the literature, many efficient algorithms exist and even \SP{larger} instances can \SP{potentially} be solved to optimality within seconds. 

Future work might look into adapting the \SP{sequence and timing of the tasks covered by a trip. }
Here, we assume a fixed sequence, however optimizing the trips as a small-sized traveling salesman problem may achieve even better results. \SP{Furthermore, the convenient but restrictive assumption that an entire trip is covered by the same mode of transport may be relaxed.}

\section*{Acknowledgements}
The authors wish to thank to anonymous referees for they valuable comments.
This work was supported by the Climate and Energy Funds (KliEn) [grant number 853767]; and the Austrian Science Fund (FWF) [P 31366].
For the purpose of open access, the author has applied a CC BY public copyright licence to any Author Accepted Manuscript version arising from this submission.


\bibliographystyle{abbrvnat}
\bibliography{manuscript}

\begin{thebibliography}{38}
\providecommand{\natexlab}[1]{#1}
\providecommand{\url}[1]{\texttt{#1}}
\expandafter\ifx\csname urlstyle\endcsname\relax
  \providecommand{\doi}[1]{doi: #1}\else
  \providecommand{\doi}{doi: \begingroup \urlstyle{rm}\Url}\fi

\bibitem[Ahuja et~al.(2001)Ahuja, Magnanti, and Orlin]{Ahuja2001}
R.~K. Ahuja, T.~L. Magnanti, and J.~B. Orlin.
\newblock Minimum cost flow problemminimum cost flow problem.
\newblock In C.~A. Floudas and P.~M. Pardalos, editors, \emph{Encyclopedia of
  Optimization}, pages 1382--1392. Springer US, Boston, MA, 2001.
\newblock ISBN 978-0-306-48332-5.
\newblock \doi{10.1007/0-306-48332-7_283}.

\bibitem[Babonneau et~al.(2006)Babonneau, du~Merle, and Vial]{Babonneau2006}
F.~Babonneau, O.~du~Merle, and J.-P. Vial.
\newblock Solving large-scale linear multicommodity flow problems with an
  active set strategy and proximal-accpm.
\newblock \emph{Operations Research}, 54\penalty0 (1):\penalty0 184--197, 2006.
\newblock \doi{10.1287/opre.1050.0262}.

\bibitem[Barnhart et~al.(1994)Barnhart, Hane, Johnson, and
  Sigismondi]{Barnhart1994}
C.~Barnhart, C.~Hane, E.~Johnson, and G.~Sigismondi.
\newblock A column generation and partitioning approach for multi-commodity
  flow problems.
\newblock \emph{Telecommunication Systems}, 3:\penalty0 239--258, 10 1994.
\newblock \doi{10.1007/BF02110307}.

\bibitem[Barnhart et~al.(2000)Barnhart, Hane, and Vance]{Barnhart2000}
C.~Barnhart, C.~A. Hane, and P.~H. Vance.
\newblock Using branch-and-price-and-cut to solve origin-destination integer
  multicommodity flow problems.
\newblock \emph{Operations Research}, 48\penalty0 (2):\penalty0 318--326, 2000.

\bibitem[Barnhart et~al.(2001)Barnhart, Krishnan, and Vance]{Barnhart2001}
C.~Barnhart, N.~Krishnan, and P.~H. Vance.
\newblock Multicommodity flow problemsmulticommodity flow problems.
\newblock In C.~A. Floudas and P.~M. Pardalos, editors, \emph{Encyclopedia of
  Optimization}, pages 1583--1591. Springer US, Boston, MA, 2001.
\newblock ISBN 978-0-306-48332-5.
\newblock \doi{10.1007/0-306-48332-7_316}.

\bibitem[bmvit(2017)]{bmvit}
bmvit.
\newblock
  \url{https://www.bmvit.gv.at/verkehr/gesamtverkehr/statistik/downloads/viz_2011_gesamtbericht_270613.pdf},
  2017.
\newblock [Online; accessed 22-August-2017].

\bibitem[Brandst{\"a}tter et~al.(2016)Brandst{\"a}tter, Gambella, Leitner,
  Malaguti, Masini, Puchinger, Ruthmair, and Vigo]{Brandstatter2016}
G.~Brandst{\"a}tter, C.~Gambella, M.~Leitner, E.~Malaguti, F.~Masini,
  J.~Puchinger, M.~Ruthmair, and D.~Vigo.
\newblock Overview of optimization problems in electric car-sharing system
  design and management.
\newblock In H.~Dawid, K.~F. Doerner, G.~Feichtinger, P.~M. Kort, and A.~Seidl,
  editors, \emph{Dynamic Perspectives on Managerial Decision Making: Essays in
  Honor of Richard F. Hartl}, pages 441--471. Springer International
  Publishing, Cham, 2016.
\newblock ISBN 978-3-319-39120-5.
\newblock \doi{10.1007/978-3-319-39120-5_24}.
\newblock URL \url{https://doi.org/10.1007/978-3-319-39120-5_24}.

\bibitem[Brandstätter et~al.(0)Brandstätter, Leitner, and
  Ljubić]{Brandstaetter2020}
G.~Brandstätter, M.~Leitner, and I.~Ljubić.
\newblock Location of charging stations in electric car sharing systems.
\newblock \emph{Transportation Science}, 0\penalty0 (0):\penalty0 null, 0.
\newblock \doi{10.1287/trsc.2019.0931}.
\newblock URL \url{https://doi.org/10.1287/trsc.2019.0931}.

\bibitem[Brunsch et~al.(2013)Brunsch, Cornelissen, Manthey, and
  R{\"o}glin]{Brunsch2013}
T.~Brunsch, K.~Cornelissen, B.~Manthey, and H.~R{\"o}glin.
\newblock Smoothed analysis of the successive shortest path algorithm.
\newblock In S.~Khanna, editor, \emph{Proceedings of the 24th ACM-SIAM
  Symposium on Discrete Algorithms}, pages 1180--1189. SIAM, 2013.
\newblock ISBN 978-1-611972-52-8.

\bibitem[Bünnagel et~al.(1998)Bünnagel, Korte, and Vygen]{Buennagel1998}
U.~Bünnagel, B.~Korte, and J.~Vygen.
\newblock Efficient implementation of the {Goldberg–Tarjan} minimum-cost flow
  algorithm.
\newblock \emph{Optimization Methods and Software}, 10\penalty0 (2):\penalty0
  157--174, 1998.
\newblock \doi{10.1080/10556789808805709}.

\bibitem[Dantzig(1963)]{Dantzig19633}
G.~Dantzig.
\newblock \emph{Linear programming and extensions}.
\newblock Rand Corporation Research Study. Princeton Univ. Press, Princeton,
  NJ, 1963.

\bibitem[de~Almeida~Correia and Antunes(2012)]{Correia2012}
G.~H. de~Almeida~Correia and A.~P. Antunes.
\newblock Optimization approach to depot location and trip selection in one-way
  carsharing systems.
\newblock \emph{Transportation Research Part E: Logistics and Transportation
  Review}, 48\penalty0 (1):\penalty0 233 -- 247, 2012.
\newblock ISSN 1366-5545.
\newblock \doi{https://doi.org/10.1016/j.tre.2011.06.003}.
\newblock Select Papers from the 19th International Symposium on Transportation
  and Traffic Theory.

\bibitem[Edmonds and Karp(1972)]{Edmonds1972}
J.~Edmonds and R.~M. Karp.
\newblock Theoretical improvements in algorithmic efficiency for network flow
  problems.
\newblock \emph{J. ACM}, 19\penalty0 (2):\penalty0 248–264, Apr. 1972.
\newblock ISSN 0004-5411.
\newblock \doi{10.1145/321694.321699}.

\bibitem[Enzi et~al.(2021{\natexlab{a}})Enzi, Parragh, Pisinger, and
  Prandtstetter]{Enzi2020}
M.~Enzi, S.~N. Parragh, D.~Pisinger, and M.~Prandtstetter.
\newblock Modeling and solving the multimodal car-and ride-sharing problem.
\newblock \emph{European Journal of Operational Research}, 293\penalty0
  (1):\penalty0 290--303, 2021{\natexlab{a}}.

\bibitem[Enzi et~al.(2021{\natexlab{b}})Enzi, Parragh, and
  Puchinger]{enzi2021bi}
M.~Enzi, S.~N. Parragh, and J.~Puchinger.
\newblock The bi-objective multimodal car-sharing problem.
\newblock \emph{OR Spectrum}, pages 1--42, 2021{\natexlab{b}}.

\bibitem[Ervolina and McCormick(1993)]{Ervolina1993}
T.~R. Ervolina and S.~McCormick.
\newblock Two strongly polynomial cut cancelling algorithms for minimum cost
  network flow.
\newblock \emph{Discrete Applied Mathematics}, 46\penalty0 (2):\penalty0 133 --
  165, 1993.
\newblock ISSN 0166-218X.
\newblock \doi{https://doi.org/10.1016/0166-218X(93)90025-J}.

\bibitem[Ford and Fulkerson(1962)]{Fulkerson1962}
L.~R. Ford and D.~R. Fulkerson.
\newblock \emph{Flows in Networks}.
\newblock Princeton University Press, 1962.
\newblock ISBN 9780691625393.

\bibitem[Goldberg and Tarjan(1989)]{Goldberg1989}
A.~Goldberg and R.~Tarjan.
\newblock Finding minimum-cost circulations by canceling negative cycles.
\newblock \emph{Journal of the ACM}, 36\penalty0 (4):\penalty0 873--886, 1
  1989.
\newblock ISSN 0004-5411.
\newblock \doi{10.1145/76359.76368}.

\bibitem[Goldberg(1997)]{Goldberg1997}
A.~V. Goldberg.
\newblock An efficient implementation of a scaling minimum-cost flow algorithm.
\newblock \emph{Journal of Algorithms}, 22\penalty0 (1):\penalty0 1 -- 29,
  1997.
\newblock ISSN 0196-6774.
\newblock \doi{https://doi.org/10.1006/jagm.1995.0805}.

\bibitem[Goldberg and Tarjan(1990)]{Goldberg1990}
A.~V. Goldberg and R.~E. Tarjan.
\newblock Finding minimum-cost circulations by successive approximation.
\newblock \emph{Mathematics of Operations Research}, 15\penalty0 (3):\penalty0
  430--466, 1990.

\bibitem[Jorge and Correia(2013)]{Jor2013}
D.~Jorge and G.~H. d.~A. Correia.
\newblock Carsharing systems demand estimation and defined operations: A
  literature review.
\newblock \emph{European Journal of Transport and Infrastructure Research},
  13:\penalty0 201--220, 03 2013.

\bibitem[Karakostas(2008)]{Karakostas2008}
G.~Karakostas.
\newblock Faster approximation schemes for fractional multicommodity flow
  problems.
\newblock \emph{ACM Transactions on Algorithms}, 4, 03 2008.
\newblock \doi{10.1145/545381.545402}.

\bibitem[Kelly et~al.(1991)Kelly, Comm, and O’Neill]{Kelly1991}
D.~Kelly, B.~R.~P. Comm, and G.~M. O’Neill.
\newblock The minimum cost flow problem and the network simplex solution
  method.
\newblock 1991.

\bibitem[Klein(1967)]{Klein1967}
M.~Klein.
\newblock {A Primal Method for Minimal Cost Flows with Applications to the
  Assignment and Transportation Problems}.
\newblock \emph{Management Science}, 14\penalty0 (3):\penalty0 205--220,
  November 1967.
\newblock \doi{10.1287/mnsc.14.3.205}.

\bibitem[Knopp et~al.(2021)Knopp, Biesinger, and Prandtstetter]{Knopp2018}
S.~Knopp, B.~Biesinger, and M.~Prandtstetter.
\newblock Mobility offer allocations in corporate settings.
\newblock \emph{EURO Journal on Computational Optimization}, 9:\penalty0
  100010, 2021.

\bibitem[Kovács(2015)]{Kovacs2015}
P.~Kovács.
\newblock Minimum-cost flow algorithms: an experimental evaluation.
\newblock \emph{Optimization Methods and Software}, 30\penalty0 (1):\penalty0
  94--127, 2015.
\newblock \doi{10.1080/10556788.2014.895828}.

\bibitem[Laporte et~al.(2018)Laporte, Meunier, and Wolfler~Calvo]{Laporte2018}
G.~Laporte, F.~Meunier, and R.~Wolfler~Calvo.
\newblock Shared mobility systems: an updated survey.
\newblock \emph{Annals of Operations Research}, 271\penalty0 (1):\penalty0
  105--126, Dec 2018.
\newblock ISSN 1572-9338.
\newblock \doi{10.1007/s10479-018-3076-8}.

\bibitem[L{\"o}bel(1996)]{Loebel1996}
A.~L{\"o}bel.
\newblock Solving large-scale real-world minimum-cost flow problems by a
  network simplex method.
\newblock Technical Report SC-96-07, ZIB, Takustr. 7, 14195 Berlin, 1996.

\bibitem[Mulley et~al.(2018)Mulley, Nelson, and Wright]{Mulley2018}
C.~Mulley, J.~D. Nelson, and S.~Wright.
\newblock Community transport meets mobility as a service: On the road to a new
  a flexible future.
\newblock \emph{Research in Transportation Economics}, 2018.
\newblock ISSN 0739-8859.
\newblock \doi{https://doi.org/10.1016/j.retrec.2018.02.004}.

\bibitem[Rechnungshof(2018)]{einkommen2018}
Rechnungshof.
\newblock Bericht des rechnungshofes: Allgemeiner einkommensbericht 2018, 2018.
\newblock URL
  \url{https://www.rechnungshof.gv.at/rh/home/home_1/home_1/Einkommensbericht_2018.pdf}.

\bibitem[R{\'e}tv{\'a}ri et~al.(2004)R{\'e}tv{\'a}ri, B{\'i}r{\'o}, and
  Cinkler]{Retvari2004}
G.~R{\'e}tv{\'a}ri, J.~B{\'i}r{\'o}, and T.~Cinkler.
\newblock A novel lagrangian-relaxation to the minimum cost multicommodity flow
  problem and its application to ospf traffic engineering.
\newblock \emph{Proceedings. ISCC 2004. Ninth International Symposium on
  Computers And Communications (IEEE Cat. No.04TH8769)}, 2:\penalty0 957--962
  Vol.2, 2004.

\bibitem[StadtWien(2017)]{stadt_wien}
StadtWien, 2017.
\newblock URL
  \url{https://www.wien.gv.at/stadtentwicklung/studien/pdf/b008404.pdf}.

\bibitem[StatistikAustria(2017)]{statstik_austria_2017}
StatistikAustria.
\newblock
  \url{http://www.statistik.at/web_de/statistiken/menschen_und_gesellschaft/arbeitsmarkt/erwerbstaetige/062875.html},
  2017.
\newblock [Online; accessed 22-August-2017].

\bibitem[Tardos(1985)]{Tardos1985}
E.~Tardos.
\newblock A strongly polynomial minimum cost circulation algorithm.
\newblock \emph{Combinatorica}, 5:\penalty0 247--256, 09 1985.
\newblock \doi{10.1007/BF02579369}.

\bibitem[Tomlin(1966)]{Tomlin1966}
J.~A. Tomlin.
\newblock Minimum-cost multicommodity network flows.
\newblock \emph{Operations Research}, 14\penalty0 (1):\penalty0 45--51, 1966.

\bibitem[Wien(2016)]{StadtWien2016}
S.~Wien.
\newblock wien.at. {W}ien - {B}ezirke im {F}okus {S}tatistiken und
  {K}ennzahlen. [online] 04 2016 [accessed: Aug 23, 2017], 2016.
\newblock https://www.wien.gv.at/statistik/pdf/bezirke-im-fokus-1-23.pdf.

\bibitem[{Wiener Linien}(2018)]{WL2018}
{Wiener Linien}.
\newblock Facts and figures, 2018.
\newblock
  https://www.wienerlinien.at/media/files/2019/betriebsangaben\_2018\_engl\_310520.pdf.

\bibitem[Zhang et~al.(2019)Zhang, Liu, and He]{Zhang2019}
D.~Zhang, Y.~Liu, and S.~He.
\newblock Vehicle assignment and relays for one-way electric car-sharing
  systems.
\newblock \emph{Transportation Research Part B: Methodological}, 120:\penalty0
  125 -- 146, 2019.
\newblock ISSN 0191-2615.
\newblock \doi{https://doi.org/10.1016/j.trb.2018.12.004}.

\end{thebibliography}

\newpage

\pagenumbering{Roman}
\setcounter{page}{1}

\appendix

\section{Appendix}

\setcounter{table}{0}
\renewcommand{\thetable}{\Alph{section}\arabic{table}}

\begin{table}[htbp]
  \centering
  \caption{Parameter value setting for the instances. The total cost are calculated as ((sloping factor * cost per km) + (sloped distance * (1 / average speed) + setup time) * cost per time + (cost of emissions * emissions per km)). \SP{The average speed of the public transportation network is the average of the average travel speeds during day time hours of bus, tram and metro as provided by \citet{WL2018} for the city of Vienna.}}
    \begin{tabular}{rl}
    \toprule
   sloping factor:  & foot: 1.1 \\
                    & bike: 1.3 \\
                    & car: 1.3 \\
                    & public transportation: 1.5 \\
                    \midrule
    CO2 emissions per km in gramm:   & foot: 0 \\
                            & bike: 0 \\
                            & combustion engine car: 200.9 \\
                            & electric car: 42.7 \\
                            & public transportation: 0 \\
                            \midrule
    cost of CO2 emissions:  & 5 euro/t \\
    \midrule
    average speed (km/h):          & foot: 5 \\
                            & bike: 16 \\
                            & car: 30 \\
                            & public transportation: 20 \\
                            \midrule
    cost per km:            & foot: 0 \\
                            & bike: 0 \\
                            & combustion engine car: 0.188 \\
                            & electric car:  0.094 \\
                            & public transportation: 0 \\
                            & taxi: 1.2 \\
   \midrule
    cost per time:                          & 19.42 euro per hour \\
     \midrule
     setup time (in minutes):               & foot: 0 \\
                                            & bike: 2 \\
                                            & car: 10 \\
                                            & public transportation: 5 \\
                                            & taxi: 5 \\
    \bottomrule
    \end{tabular}%
  \label{tab:instance-data}%
\end{table}%

\begin{table}[htbp]
  \centering
  \caption{Comparison of total cost split into combustion engine cars (\car) and other MOTs, and savings for increasing number of $u$ and $m$ for \minflowcar. Share of total cost of the respective car and MOT costs given in '\car / total' and 'other MOTs / total'.}
    \begin{tabular}{ccccccccc}
    \toprule
    \multicolumn{1}{l}{\textit{ u }} & \multicolumn{1}{l}{\textit{ m }} & \multicolumn{1}{l}{ \car } & \multicolumn{1}{l}{ other MOTs } & \multicolumn{1}{l}{ total } & \multicolumn{1}{l}{ savings } &       & \multicolumn{1}{l}{ \car / total } & \multicolumn{1}{l}{ other MOTs / total } \\
    \midrule
    \multirow{4}[2]{*}{    20 } &      4  &                  155  &                  401  &                  555  &                  49  &       &                0.28  &                0.72  \\
          &      8  &                  243  &                  294  &                  537  &                  68  &       &                0.45  &                0.55  \\
          &   20  &                  316  &                  204  &                  520  &                  85  &       &                0.61  &                0.39  \\
          &   40  &                  318  &                  202  &                  520  &                  85  &       &                0.61  &                0.39  \\
    \midrule
    \multirow{4}[2]{*}{    50 } &      4  &                  264  &              1,423  &              1,686  &                122  &       &                0.16  &                0.84  \\
          &      8  &                  467  &              1,160  &              1,626  &                182  &       &                0.29  &                0.71  \\
          &   20  &                  846  &                  704  &              1,550  &                258  &       &                0.55  &                0.45  \\
          &   40  &              1,050  &                  482  &              1,532  &                276  &       &                0.69  &                0.31  \\
    \midrule
    \multirow{4}[2]{*}{  100 } &      4  &                  291  &              3,090  &              3,381  &                178  &       &                0.09  &                0.91  \\
          &      8  &                  553  &              2,735  &              3,288  &                270  &       &                0.17  &                0.83  \\
          &   20  &              1,155  &              2,008  &              3,163  &                395  &       &                0.37  &                0.63  \\
          &   40  &              1,820  &              1,289  &              3,109  &                449  &       &                0.59  &                0.41  \\
    \midrule
    \multirow{4}[2]{*}{  150 } &      4  &                  320  &              5,146  &              5,466  &                221  &       &                0.06  &                0.94  \\
          &      8  &                  575  &              4,752  &              5,327  &                360  &       &                0.11  &                0.89  \\
          &   20  &              1,235  &              3,832  &              5,067  &                619  &       &                0.24  &                0.76  \\
          &   40  &              2,079  &              2,797  &              4,875  &                811  &       &                0.43  &                0.57  \\
    \midrule
    \multirow{4}[2]{*}{  200 } &      4  &                  338  &              7,087  &              7,424  &                241  &       &                0.05  &                0.95  \\
          &      8  &                  624  &              6,635  &              7,260  &                406  &       &                0.09  &                0.91  \\
          &   20  &              1,320  &              5,608  &              6,928  &                738  &       &                0.19  &                0.81  \\
          &   40  &              2,310  &              4,339  &              6,648  &            1,017  &       &                0.35  &                0.65  \\
    \midrule
    \multirow{4}[2]{*}{  250 } &      4  &                  373  &              9,063  &              9,436  &                289  &       &                0.04  &                0.96  \\
          &      8  &                  670  &              8,567  &              9,238  &                487  &       &                0.07  &                0.93  \\
          &   20  &              1,470  &              7,391  &              8,861  &                864  &       &                0.17  &                0.83  \\
          &   40  &              2,506  &              6,048  &              8,555  &            1,170  &       &                0.29  &                0.71  \\
    \midrule
    \multirow{4}[2]{*}{  300 } &      4  &                  350  &              9,966  &            10,316  &                247  &       &                0.03  &                0.97  \\
          &      8  &                  638  &              9,486  &            10,124  &                439  &       &                0.06  &                0.94  \\
          &   20  &              1,427  &              8,348  &              9,775  &                789  &       &                0.15  &                0.85  \\
          &   40  &              2,495  &              7,017  &              9,512  &            1,051  &       &                0.26  &                0.74  \\
    \bottomrule
    \end{tabular}%
  \label{tab:appendix:cost-car-percentage-nopref}%
\end{table}%

\begin{table}[htbp]
  \centering
  \caption{Comparison of total cost split into electric cars (\ecar) and other MOTs, and savings for increasing number of $u$ and $m$ for \minflowecar. Share of total cost of the respective car and MOT costs given in '\ecar / total' and 'other MOTs / total'.}
    \begin{tabular}{crrrrrrrr}
    \toprule
    \multicolumn{1}{l}{\textit{ u }} & \multicolumn{1}{l}{\textit{ m }} & \multicolumn{1}{l}{ \ecar } & \multicolumn{1}{l}{ other MOTs } & \multicolumn{1}{l}{ total } & \multicolumn{1}{l}{ savings } &       & \multicolumn{1}{l}{ \ecar / total } & \multicolumn{1}{l}{ other MOTs / total } \\
    \midrule
    \multirow{4}[2]{*}{                20 } &                   4  &               162  &               379  &               541  &               64  &       &              0.30  &              0.70  \\
          &                   8  &               271  &               243  &               514  &               91  &       &              0.53  &              0.47  \\
          &                 20  &               348  &               143  &               491  &             114  &       &              0.71  &              0.29  \\
          &                 40  &               352  &               138  &               491  &             114  &       &              0.72  &              0.28  \\
    \midrule
    \multirow{4}[2]{*}{                50 } &                   4  &               243  &            1,420  &            1,663  &             145  &       &              0.15  &              0.85  \\
          &                   8  &               430  &            1,155  &            1,588  &             220  &       &              0.27  &              0.73  \\
          &                 20  &               850  &               625  &            1,475  &             333  &       &              0.58  &              0.42  \\
          &                 40  &            1,098  &               340  &            1,438  &             370  &       &              0.76  &              0.24  \\
    \midrule
    \multirow{4}[2]{*}{              100 } &                   4  &               266  &            3,089  &            3,355  &             204  &       &              0.08  &              0.92  \\
          &                   8  &               509  &            2,731  &            3,239  &             319  &       &              0.16  &              0.84  \\
          &                 20  &            1,084  &            1,977  &            3,062  &             497  &       &              0.35  &              0.65  \\
          &                 40  &            1,776  &            1,175  &            2,951  &             608  &       &              0.60  &              0.40  \\
    \midrule
    \multirow{4}[2]{*}{              150 } &                   4  &               298  &            5,139  &            5,437  &             250  &       &              0.05  &              0.95  \\
          &                   8  &               539  &            4,736  &            5,275  &             412  &       &              0.10  &              0.90  \\
          &                 20  &            1,150  &            3,808  &            4,957  &             729  &       &              0.23  &              0.77  \\
          &                 40  &            1,960  &            2,733  &            4,693  &             993  &       &              0.42  &              0.58  \\
    \midrule
    \multirow{4}[2]{*}{              200 } &                   4  &               310  &            7,083  &            7,393  &             272  &       &              0.04  &              0.96  \\
          &                   8  &               577  &            6,625  &            7,202  &             463  &       &              0.08  &              0.92  \\
          &                 20  &            1,237  &            5,572  &            6,809  &             857  &       &              0.18  &              0.82  \\
          &                 40  &            2,165  &            4,279  &            6,444  &          1,222  &       &              0.34  &              0.66  \\
    \midrule
    \multirow{4}[2]{*}{              250 } &                   4  &               351  &            9,050  &            9,401  &             324  &       &              0.04  &              0.96  \\
          &                   8  &               625  &            8,551  &            9,176  &             549  &       &              0.07  &              0.93  \\
          &                 20  &            1,371  &            7,357  &            8,728  &             997  &       &              0.16  &              0.84  \\
          &                 40  &            2,338  &            5,994  &            8,332  &          1,393  &       &              0.28  &              0.72  \\
    \midrule
    \multirow{4}[2]{*}{              300 } &                   4  &               322  &            9,962  &          10,284  &             280  &       &              0.03  &              0.97  \\
          &                   8  &               596  &            9,470  &          10,066  &             497  &       &              0.06  &              0.94  \\
          &                 20  &            1,332  &            8,315  &            9,646  &             917  &       &              0.14  &              0.86  \\
          &                 40  &            2,339  &            6,952  &            9,291  &          1,272  &       &              0.25  &              0.75  \\
    \bottomrule
    \end{tabular}%
  \label{tab:appendix:cost-ecar-percentage-nopref}%
\end{table}%

\begin{table}[htbp]
  \centering
  \caption{Comparison of total cost split into combustion engine cars (\car), electric cars (\ecar) and other MOTs, and savings for increasing number of $u$ and $m$ for \multcomm. Share of total cost of the respective car and MOT costs given in 'car-type / total' and 'other MOTs / total'.}
   \setlength{\tabcolsep}{2pt}
    \begin{tabular}{crrrrrrrrrr}
    \toprule
    \multicolumn{1}{l}{\textit{ u =}} & \multicolumn{1}{l}{\textit{ m }} & \multicolumn{1}{l}{ \car } & \multicolumn{1}{l}{\ecar} & \multicolumn{1}{l}{ other MOTs } & \multicolumn{1}{l}{ total } & \multicolumn{1}{l}{ savings } &       & \multicolumn{1}{l}{ \car / total } & \multicolumn{1}{l}{ \ecar / total } & \multicolumn{1}{l}{ other MOTs / total } \\
    \midrule
    \multirow{4}[2]{*}{    20 } &      4  &                 64  &                 91  &               392  &               547  &               58  &       &              0.12  &              0.17  &              0.72  \\
          &      8  &                 88  &               163  &               272  &               522  &               82  &       &              0.17  &              0.31  &              0.52  \\
          &    20  &                 48  &               298  &               149  &               495  &             110  &       &              0.10  &              0.60  &              0.30  \\
          &    40  &                   2  &               347  &               142  &               491  &             114  &       &              0.00  &              0.71  &              0.29  \\
    \midrule
    \multirow{4}[2]{*}{    50 } &      4  &               112  &               139  &            1,421  &            1,673  &             135  &       &              0.07  &              0.08  &              0.85  \\
          &      8  &               194  &               249  &            1,159  &            1,603  &             205  &       &              0.12  &              0.16  &              0.72  \\
          &    20  &               270  &               536  &               694  &            1,500  &             308  &       &              0.18  &              0.36  &              0.46  \\
          &    40  &               196  &               859  &               401  &            1,456  &             352  &       &              0.13  &              0.59  &              0.28  \\
    \midrule
    \multirow{4}[2]{*}{  100 } &      4  &               124  &               152  &            3,090  &            3,366  &             193  &       &              0.04  &              0.05  &              0.92  \\
          &      8  &               235  &               291  &            2,734  &            3,260  &             299  &       &              0.07  &              0.09  &              0.84  \\
          &    20  &               468  &               637  &            1,998  &            3,102  &             456  &       &              0.15  &              0.21  &              0.64  \\
          &    40  &               634  &            1,102  &            1,270  &            3,006  &             552  &       &              0.21  &              0.37  &              0.42  \\
    \midrule
    \multirow{4}[2]{*}{  150 } &      4  &               142  &               161  &            5,146  &            5,450  &             237  &       &              0.03  &              0.03  &              0.94  \\
          &      8  &               234  &               314  &            4,748  &            5,296  &             391  &       &              0.04  &              0.06  &              0.90  \\
          &    20  &               490  &               691  &            3,820  &            5,000  &             686  &       &              0.10  &              0.14  &              0.76  \\
          &    40  &               800  &            1,191  &            2,771  &            4,762  &             925  &       &              0.17  &              0.25  &              0.58  \\
    \midrule
    \multirow{4}[2]{*}{  200 } &      4  &               155  &               168  &            7,085  &            7,408  &             258  &       &              0.02  &              0.02  &              0.96  \\
          &      8  &               267  &               331  &            6,628  &            7,226  &             439  &       &              0.04  &              0.05  &              0.92  \\
          &    20  &               530  &               740  &            5,586  &            6,856  &             810  &       &              0.08  &              0.11  &              0.81  \\
          &    40  &               902  &            1,314  &            4,306  &            6,522  &          1,144  &       &              0.14  &              0.20  &              0.66  \\
    \midrule
    \multirow{4}[2]{*}{  250 } &      4  &               161  &               193  &            9,063  &            9,417  &             308  &       &              0.02  &              0.02  &              0.96  \\
          &      8  &               281  &               363  &            8,558  &            9,202  &             523  &       &              0.03  &              0.04  &              0.93  \\
          &    20  &               587  &               818  &            7,376  &            8,781  &             944  &       &              0.07  &              0.09  &              0.84  \\
          &    40  &               956  &            1,442  &            6,018  &            8,416  &          1,309  &       &              0.11  &              0.17  &              0.72  \\
    \midrule
    \multirow{4}[2]{*}{  300 } &      4  &               154  &               179  &            9,964  &          10,298  &             265  &       &              0.01  &              0.02  &              0.97  \\
          &      8  &               271  &               337  &            9,482  &          10,091  &             473  &       &              0.03  &              0.03  &              0.94  \\
          &    20  &               570  &               790  &            8,338  &            9,698  &             866  &       &              0.06  &              0.08  &              0.86  \\
          &    40  &               969  &            1,424  &            6,983  &            9,376  &          1,187  &       &              0.10  &              0.15  &              0.74  \\
    \bottomrule
    \end{tabular}%
  \label{tab:appendix:cost-multcomm-percentage-nopref}%
\end{table}%

\begin{table}[htbp]
  \centering
  \caption{Solution time in seconds for \minflowcar, \minflowecar, \multcomm\ for an increasing number of $u$ and $m$.}
    \begin{tabular}{crccc}
    \toprule
    \multicolumn{1}{l}{\textit{u}} & \multicolumn{1}{l}{\textit{m}} & \multicolumn{1}{c}{ \minflowecar} & \minflowcar     & \multcomm \\
    \midrule
    \multirow{4}[1]{*}{20} & 4     & 0.0   &     0.0  &     0.1  \\
          & 8     & 0.0   &     0.0  &     0.1  \\
          & 20    & 0.0   &     0.0  &     0.0  \\
          & 40    & 0.0   &     0.0  &     0.1  \\
          &       &       &       &  \\
    \multirow{4}[0]{*}{50} & 4     & 0.1   &     0.2  &     0.4  \\
          & 8     & 0.1   &     0.2  &     0.4  \\
          & 20    & 0.2   &     0.2  &     0.4  \\
          & 40    & 0.1   &     0.2  &     0.4  \\
          &       &       &       &  \\
    \multirow{4}[0]{*}{100} & 4     & 0.7   &     0.7  &     1.6  \\
          & 8     & 0.7   &     0.7  &     1.6  \\
          & 20    & 0.7   &     0.7  &     1.6  \\
          & 40    & 0.7   &     0.7  &     1.6  \\
          &       &       &       &  \\
    \multirow{4}[0]{*}{150} & 4     & 1.6   &     1.6  &     3.7  \\
          & 8     & 1.6   &     1.6  &     3.6  \\
          & 20    & 1.6   &     1.6  &     3.7  \\
          & 40    & 1.6   &     1.6  &     3.6  \\
          &       &       &       &  \\
    \multirow{4}[0]{*}{200} & 4     & 3.0   &     3.1  &     7.0  \\
          & 8     & 3.1   &     3.1  &     6.8  \\
          & 20    & 3.1   &     3.1  &     6.7  \\
          & 40    & 3.0   &     3.0  &     6.9  \\
          &       &       &       &  \\
    \multirow{4}[0]{*}{250} & 4     & 5.0   &     4.9  &   11.0  \\
          & 8     & 4.8   &     5.0  &   10.8  \\
          & 20    & 4.9   &     4.9  &   11.0  \\
          & 40    & 4.8   &     5.0  &   10.9  \\
          &       &       &       &  \\
    \multirow{4}[1]{*}{300} & 4     & 7.3   &     7.4  &   16.3  \\
          & 8     & 7.2   &     7.6  &   17.1  \\
          & 20    & 7.3   &     7.4  &   16.8  \\
          & 40    & 7.4   &     7.4  &   16.8  \\
    \bottomrule
    \end{tabular}%
  \label{tab:time-nopref}%
\end{table}%

\begin{table}[htbp]
  \centering
  \caption{Total cost comparison split for an increasing $u$ and averaged over all $m$ for \minflowcar, \minflowecar, and \multcomm. Column 'cost' gives the absolute cost of the respective model, 'comp.' compares the cost to \minflowecar\ where it is set as (cost of the model / cost of \minflowcar).}
    \begin{tabular}{cccccc}
    \toprule
          & \multicolumn{1}{c}{ \minflowecar } & \multicolumn{2}{c}{ \multcomm} & \multicolumn{2}{c}{ \minflowcar } \\
    \multicolumn{1}{l}{ u } & \multicolumn{1}{c}{ cost } & \multicolumn{1}{c}{ cost } & \multicolumn{1}{c}{ comp. } & \multicolumn{1}{c}{ cost } & \multicolumn{1}{c}{ comp. } \\
    \midrule
                    20  &                    509  &               514  &                 1.01  &               533  &              1.05  \\
                    50  &                 1,541  &            1,558  &                 1.01  &            1,599  &              1.04  \\
                  100  &                 3,152  &            3,184  &                 1.01  &            3,235  &              1.03  \\
                  150  &                 5,091  &            5,127  &                 1.01  &            5,184  &              1.02  \\
                  200  &                 6,962  &            7,003  &                 1.01  &            7,065  &              1.01  \\
                  250  &                 8,909  &            8,954  &                 1.00  &            9,022  &              1.01  \\
                  300  &                 9,822  &            9,866  &                 1.00  &            9,932  &              1.01  \\
    \bottomrule
    \end{tabular}%
  \label{tab:comparison-models}%
\end{table}%

\begin{table}[htbp]
  \centering
  \caption{Average cost for combustion engine cars (\car) and other MOTs, in total, and average savings for \minflowcar\ and the different preference variants (prefVar0-prefVar6). The values are given for an increasing number of $u$ and they are averaged over $m = 4, 8, 20, 40$.}
    \begin{tabular}{lrrrrrrr}
    \toprule
    u =    & 20    & 50    & 100   & 150   & 200   & 250   & 300 \\
    \midrule
          & \multicolumn{7}{c}{\minflowcar} \\
     \car  & 258 &                 656  &                 955  &             1,052  &              1,148  &             1,255  &                 1,227  \\
     other MOTs  & 275 &                 942  &             2,281  &             4,132  &             5,917  &            7,767  &                8,704  \\
     total  &               533  &             1,599  &            3,235  &             5,184  &            7,065  &            9,022  &                9,932  \\
     savings  &               72  &              209  &              323  &              503  &               601  &              703  &                   631  \\
    \midrule
          & \multicolumn{7}{c}{prefVar0} \\
     \car  & 180 &                 429  &                 624  &                 783  &                 825  &                 902  &                     872  \\
     other MOTs  & 424 &             1,495  &             3,361  &            5,930  &            8,302  &          10,854  &              12,093  \\
     total  &               604  &             1,925  &            3,985  &             6,713  &             9,127  &           11,757  &              12,964  \\
     savings  &             215  &              595  &             1,119  &           1,682  &           1,939  &           2,109  &              2,264  \\
    \midrule
          & \multicolumn{7}{c}{prefVar1}   \\
     \car  &               237  &                 580  &                 873  &                 989  &             1,080  &              1,162  &                  1,169  \\
     other MOTs  &               323  &             1,084  &            2,475  &            4,358  &             6,198  &             8,154  &                 9,014  \\
     total  &               560  &             1,664  &            3,348  &            5,347  &            7,279  &             9,316  &               10,184  \\
     savings  &               76  &              209  &              343  &              590  &              698  &               818  &                  787  \\
    \midrule
          & \multicolumn{7}{c}{ prefVar2 }   \\
     \car  &                  42  &                 200  &                 359  &                  510  &                  618  &                 726  &                     722  \\
     other MOTs  &               558  &             1,570  &             3,150  &            5,047  &            6,893  &            8,783  &                9,669  \\
     total  &               600  &             1,770  &            3,509  &            5,558  &              7,511  &            9,509  &               10,391  \\
     savings  &                 11  &                 56  &                 74  &               196  &               213  &               310  &                  252  \\
    \midrule
          & \multicolumn{7}{c}{ prefVar3 }   \\
     \car  &                144  &                 425  &                 707  &                 848  &                 943  &             1,036  &                 1,030  \\
     other MOTs  &               437  &              1,281  &            2,698  &            4,569  &             6,415  &             8,331  &                 9,231  \\
     total  &                581  &             1,706  &            3,405  &             5,417  &            7,358  &            9,367  &               10,261  \\
     savings  &               39  &               149  &              245  &              444  &               471  &              642  &                  522  \\
    \midrule
          & \multicolumn{7}{c}{ prefVar4 }  \\
     \car  &               245  &                 605  &                  901  &             1,003  &               1,101  &              1,187  &                  1,188  \\
     other MOTs  &               325  &             1,084  &            2,496  &            4,475  &            6,327  &            8,375  &                9,220  \\
     total  &                571  &             1,689  &            3,397  &            5,478  &            7,429  &            9,562  &              10,408  \\
     savings  &               84  &               251  &              409  &               713  &                811  &               981  &                   910  \\
    \midrule
          & \multicolumn{7}{c}{ prefVar5 } \\
     \car  &                  42  &                 209  &                 376  &                 522  &                 650  &                 754  &                     747  \\
     other MOTs  &               559  &             1,565  &             3,142  &            5,053  &            6,882  &            8,788  &                9,682  \\
     total  &                601  &             1,774  &             3,518  &            5,575  &            7,532  &            9,542  &              10,429  \\
     savings  &                 11  &                 63  &                 84  &              233  &              255  &              376  &                  306  \\
    \midrule
          & \multicolumn{7}{c}{ prefVar6 }  \\
     \car  &                148  &                 442  &                 732  &                 868  &                 963  &             1,059  &                 1,045  \\
     other MOTs  &               439  &             1,278  &            2,700  &            4,622  &            6,474  &            8,445  &                9,337  \\
     total  &               587  &             1,720  &            3,432  &            5,489  &            7,437  &            9,504  &              10,382  \\
     savings  &               45  &               170  &              278  &              533  &               551  &              755  &                   615  \\
    \bottomrule
    \end{tabular}%
  \label{tab:av-prefVar-car}%
\end{table}%

\begin{table}[htbp]
  \centering
  \caption{Average cost for one type of vehicle (combustion engine cars, \car), \ecar\ and other MOTs, in total and average savings for \multcomm and the different preference variants (prefVar0-prefVar6). The values are given for an increasing number of $u$ and averages over all $m$.}
  \small
     \begin{tabular}{lrrrrrrr}
    \toprule
    u =     & 20    &                     50    &                   100    &                   150    &                  200    &                  250    &                      300    \\
    \midrule
          & \multicolumn{7}{c}{\multcomm} \\
    \car & 50    & 193   & 365   & 417   & 463   & 496   & 491 \\
    \ecar & 225   & 446   & 545   & 589   & 638   & 704   & 683 \\
    other MOTs & 239   & 919   & 2273  & 4121  & 5901  & 7754  & 8692 \\
    total & 514   & 1558  & 3184  & 5127  & 7003  & 8954  & 9866 \\
    savings & 91   & 250  & 375  & 560  & 663  & 771  & 698 \\
    \midrule
          & \multicolumn{7}{c}{preVar0} \\
    \car & 27    & 86    & 163   & 303   & 304   & 343   & 339 \\
    \ecar & 147   & 295   & 387   & 455   & 477   & 518   & 505 \\
    other MOTs & 418   & 1520  & 3401  & 5916  & 8303  & 10848 & 12075 \\
    total & 593   & 1900  & 3951  & 6673  & 9084  & 11709 & 12919 \\
    savings & 226  & 620  & 1153 & 1722 & 1982 & 2156 & 2309 \\
    \midrule
          & \multicolumn{7}{c}{preVar1} \\
    \car & 110   & 253   & 413   & 471   & 510   & 534   & 558 \\
    \ecar & 130   & 330   & 440   & 486   & 535   & 588   & 573 \\
    other MOTs & 313   & 1058  & 2461  & 4354  & 6193  & 8150  & 9010 \\
    total & 553   & 1641  & 3314  & 5311  & 7238  & 9272  & 10141 \\
    savings & 83   & 232  & 377  & 626  & 738  & 862  & 829 \\
    \midrule
          & \multicolumn{7}{c}{prefVar2} \\
    \car & 31    & 96    & 172   & 254   & 289   & 336   & 348 \\
    \ecar & 13    & 114   & 191   & 246   & 324   & 376   & 371 \\
    other MOTs & 555   & 1553  & 3133  & 5040  & 6874  & 8768  & 9646 \\
    total & 599   & 1763  & 3496  & 5539  & 7488  & 9481  & 10365 \\
    savings & 12   & 63   & 88   & 214  & 236  & 338  & 278 \\
    \midrule
          & \multicolumn{7}{c}{prefVar3} \\
    \car & 64    & 180   & 330   & 398   & 451   & 483   & 491 \\
    \ecar & 80    & 257   & 374   & 432   & 466   & 524   & 505 \\
    other MOTs & 432   & 1252  & 2675  & 4554  & 6404  & 8323  & 9227 \\
    total & 576   & 1689  & 3379  & 5384  & 7321  & 9329  & 10222 \\
    savings & 44   & 165  & 271  & 477  & 508  & 680  & 560 \\
    \midrule
          & \multicolumn{7}{c}{prefVar4} \\
    \car & 139   & 332   & 454   & 516   & 543   & 596   & 593 \\
    \ecar & 106   & 270   & 447   & 487   & 558   & 592   & 595 \\
    other MOTs & 325   & 1087  & 2496  & 4475  & 6327  & 8374  & 9220 \\
    total & 571   & 1689  & 3397  & 5478  & 7429  & 9562  & 10408 \\
    savings & 84   & 251  & 409  & 713  & 811  & 981  & 910 \\
    \midrule
          & \multicolumn{7}{c}{prefVar5} \\
    \car & 41    & 139   & 260   & 323   & 363   & 399   & 383 \\
    \ecar & 1     & 75    & 116   & 194   & 277   & 348   & 361 \\
    other MOTs & 559   & 1560  & 3142  & 5057  & 6892  & 8795  & 9685 \\
    total & 601   & 1774  & 3518  & 5575  & 7532  & 9542  & 10429 \\
    savings & 11   & 63   & 84   & 233  & 255  & 376  & 306 \\
    \midrule
          & \multicolumn{7}{c}{prefVar6} \\
    \car & 88    & 253   & 382   & 439   & 482   & 527   & 515 \\
    \ecar & 59    & 192   & 353   & 428   & 481   & 532   & 531 \\
    other MOTs & 439   & 1274  & 2698  & 4623  & 6473  & 8445  & 9337 \\
    total & 587   & 1720  & 3432  & 5489  & 7437  & 9504  & 10382 \\
    savings & 45   & 170  & 278  & 533  & 551  & 755  & 615 \\
    \bottomrule
    \end{tabular}%
  \label{tab:av-prefVar-multcomm}%
\end{table}%

\begin{table}[htbp]
  \centering
  \caption{Total cost comparison for OF:time split into combustion engine cars (\car) and other MOTs for an increasing number of $u$ and $m$ for \minflowcar.}
    \begin{tabular}{clrrrrrrr}
    \toprule
          & \textit{ u =} &                             20  &                             50  &                        100  &                          150  &                           200  &                           250  &                           300  \\
    \midrule
    \multirow{3}[2]{*}{      4 } &  \car  &                          186  &                          275  &                        293  &                          323  &                           341  &                           374  &                           355  \\
          &  other MOTs  &                          383  &                      1,455  &                    3,176  &                      5,473  &                       7,479  &                       9,691  &                    10,380  \\
          &  total  &                          569  &                      1,730  &                    3,469  &                      5,796  &                       7,820  &                    10,065  &                    10,735  \\
    \midrule
    \multirow{3}[2]{*}{      8 } &  \car  &                          310  &                          482  &                        560  &                          593  &                           640  &                           690  &                           665  \\
          &  other MOTs  &                          235  &                      1,166  &                    2,774  &                      4,980  &                       6,922  &                       9,047  &                       9,751  \\
          &  total  &                          545  &                      1,648  &                    3,334  &                      5,573  &                       7,562  &                       9,738  &                    10,416  \\
    \midrule
    \multirow{3}[2]{*}{   20 } &  \car  &                          418  &                          954  &                    1,192  &                      1,279  &                       1,370  &                       1,482  &                       1,482  \\
          &  other MOTs  &                          108  &                          602  &                    1,973  &                      3,891  &                       5,702  &                       7,656  &                       8,405  \\
          &  total  &                          526  &                      1,556  &                    3,165  &                      5,169  &                       7,071  &                       9,138  &                       9,887  \\
    \midrule
    \multirow{3}[2]{*}{   40 } &  \car  &                          418  &                      1,277  &                    1,971  &                      2,161  &                       2,389  &                       2,585  &                       2,605  \\
          &  other MOTs  &                          107  &                          268  &                    1,145  &                      2,738  &                       4,298  &                       6,127  &                       6,947  \\
          &  total  &                          526  &                      1,545  &                    3,116  &                      4,899  &                       6,687  &                       8,712  &                       9,552  \\
    \bottomrule
    \end{tabular}%
  \label{tab:timeOF-car}%
\end{table}%

\begin{table}[htbp]
  \centering
  \caption{Comparison of total cost for OF:time split into \car, \ecar (= combustion engine and electric cars) and other MOTs for an increasing number of $u$ and $m$ for \multcomm.}
    \begin{tabular}{clrrrrrrr}
    \toprule
          & \textit{ u =} &                             20  &                             50  &                        100  &                          150  &                           200  &                           250  &                           300  \\
    \midrule
    \multirow{4}[2]{*}{      4 } &  \car  &                             98  &                          140  &                        144  &                          157  &                           174  &                           195  &                           184  \\
          &  \ecar  &                             79  &                          123  &                        135  &                          151  &                           152  &                           166  &                           156  \\
          &  other MOTs  &                          385  &                      1,455  &                    3,176  &                      5,473  &                       7,479  &                       9,689  &                    10,379  \\
          &  total  &                          562  &                      1,718  &                    3,455  &                      5,781  &                       7,805  &                    10,049  &                    10,719  \\
    \midrule
    \multirow{4}[2]{*}{      8 } &  \car  &                          158  &                          249  &                        274  &                          284  &                           318  &                           357  &                           321  \\
          &  \ecar  &                          138  &                          212  &                        262  &                          283  &                           293  &                           301  &                           309  \\
          &  other MOTs  &                          236  &                      1,167  &                    2,773  &                      4,978  &                       6,922  &                       9,049  &                       9,755  \\
          &  total  &                          532  &                      1,628  &                    3,309  &                      5,546  &                       7,533  &                       9,708  &                    10,385  \\
    \midrule
    \multirow{4}[2]{*}{   20 } &  \car  &                          204  &                          482  &                        605  &                          643  &                           694  &                           745  &                           743  \\
          &  \ecar  &                          190  &                          433  &                        537  &                          581  &                           620  &                           673  &                           676  \\
          &  other MOTs  &                          114  &                          602  &                    1,973  &                      3,890  &                       5,698  &                       7,655  &                       8,403  \\
          &  total  &                          508  &                      1,517  &                    3,115  &                      5,114  &                       7,012  &                       9,073  &                       9,822  \\
    \midrule
    \multirow{4}[2]{*}{   40 } &  \car  &                          248  &                          689  &                    1,003  &                      1,072  &                       1,224  &                       1,305  &                       1,296  \\
          &  \ecar  &                          156  &                          537  &                        886  &                          995  &                       1,066  &                       1,170  &                       1,197  \\
          &  other MOTs  &                          107  &                          271  &                    1,146  &                      2,740  &                       4,296  &                       6,125  &                       6,946  \\
          &  total  &                          511  &                      1,497  &                    3,035  &                      4,806  &                       6,587  &                       8,600  &                       9,439  \\
    \bottomrule
    \end{tabular}%
  \label{tab:timeOF-multcomm}%
\end{table}%

\end{document}